\documentclass[10pt,final,journal,letterpaper,oneside,twocolumn]{IEEEtran}
\usepackage{amssymb}
\usepackage{algorithm}

\usepackage{color}
\usepackage[dvipsnames]{xcolor}
\usepackage{algpseudocode}
\usepackage{cite}
\usepackage{graphicx}

\usepackage{amsfonts}
\usepackage[cmex10]{amsmath}
\ifCLASSOPTIONcompsoc
\usepackage[tight,normalsize,sf,SF]{subfigure}
\else
\usepackage[tight,footnotesize]{subfigure}
\fi

\usepackage{amsmath}
\usepackage{epstopdf}
\usepackage{bbm}
\usepackage{array}
\usepackage{mdwmath}
\usepackage{mdwtab}
\usepackage{multirow}

\newenvironment{remark}[1][Remark]{\begin{trivlist}
\item[\hskip \labelsep {\bfseries #1}]}{\end{trivlist}}

\begin{document}

\title{Markov Decision Model for Adaptive Scheduling of Stored Scalable Videos}

\author{Chao Chen,~\IEEEmembership{Student Member,~IEEE,}
        Robert W. Heath Jr.,~\IEEEmembership{Fellow,~IEEE,}\\
        Alan C. Bovik,~\IEEEmembership{Fellow,~IEEE,}
        and~Gustavo~de~Veciana,~\IEEEmembership{Fellow,~IEEE}% <-this % stops a space
\thanks{The authors are with Department of Electrical and Computer Engineering,
The University of Texas at Austin, 1 University Station C0803, Austin TX - 78712-0240, USA. E-mail: chao.chen@utexas.edu, rheath@ece.utexas.edu, bovik@ece.utexas.edu, gustavo@ece.utexas.edu. This research was supported in part by Intel Inc. and Cisco Corp. under the VAWN program.
Copyright (c) 2013 IEEE. Personal use of this material is permitted. However, permission to use this material for any other purposes must be obtained from the IEEE by sending an email to pubs-permissions@ieee.org.}

}

\maketitle

\begin{abstract}
\textcolor{black}{We propose two scheduling algorithms that seek to optimize the quality of scalably coded videos that have been stored at a video server before transmission.} The first scheduling algorithm is derived from a Markov Decision Process (MDP) formulation developed here. We model the dynamics of the channel as a Markov chain and reduce the problem of dynamic video scheduling to a tractable Markov decision problem over a finite state space. Based on the MDP formulation, a near-optimal scheduling policy is computed that minimize the mean square error. Using insights taken from the development of the optimal MDP-based scheduling policy, the second proposed scheduling algorithm is an online scheduling method that only requires easily measurable knowledge of the channel dynamics, and is thus viable in practice. Simulation results show that the performance of both scheduling algorithms is close to a performance upper bound also derived in this paper.
\end{abstract}


% For peer review papers, you can put extra information on the cover
% page as needed:
% \ifCLASSOPTIONpeerreview
% \begin{center} \bfseries EDICS Category: 3-BBND \end{center}
% \fi
%
% For peerreview papers, this IEEEtran command inserts a page break and
% creates the second title. It will be ignored for other modes.
\IEEEpeerreviewmaketitle

\section{Introduction}
\label{sec:intro}
\IEEEPARstart{T}{he} variation of wireless channel capacity and tight delay constraints make the delivery of video difficult. Although adaptive transmission strategies such as adaptive video data scheduling can be employed, deriving the optimal adaptive transmission policy is difficult because the transmission strategies taken at different time are coupled with each other via receiver buffer state. Furthermore, due to the nature of predictive video coding algorithms, a video frame can be decoded only when its predictors have been received. Hence, the prediction structure of the video codec enforces a partial order on the transmissions of the video packets, which limits the flexibility of adaptive video transmission.

Scalable video coding (SVC) is one approach to enable flexible video transmission over channels with varying throughput \cite{SVC,FGS}. An SVC video encoder produces a layered video stream that contains a base layer and several enhancement layers. If the throughput is low, the transmitter can choose to transmit the base layer only, which provides a moderate, but acceptable, degree of visual quality at the receiver. If the channel conditions improve, the transmitter can transmit one, or more, enhancement layers to further improve the visual quality. Conceptually, SVC provides a means to adapt the data rate for wireless video transmission. The wireless transmitter can adapt the data rate by selectively scheduling video data associated with various layers for transmission rather than transcoding the video sequence into a different rate.

Designing scalable video scheduling algorithms for wireless channels is a complex task. The scheduling policy depends, not only on the channel conditions, but also, on the receiver buffer state. For example, if the receiver has successfully buffered base layer data over many frames, the scheduler could choose to transmit some enhancement layer data to improve the video quality even if the throughput is low. At any time, the scheduling decision will determine the receiver buffer state which, in turn, affects the future scheduling decisions. Therefore, adaptive video data scheduling is a sequential decision problem. The most natural way to address such problems is to model the dynamics of the channel as a finite state Markov chain and to employ a Markov decision process (MDP)-based formulation to study scheduling methods. \textcolor{black}{For stored video transmission, however, directly determining an optimal scheduling policy using an MDP formulation is not possible, because the system state space is infinitely large} (see Section. \ref{sec:simplestate}). Moreover, in a practical wireless network, a model for the dynamics of the channel states is not typically available, which limits the applicability of this approach.

\subsection{Contributions}
The objective of this paper is to leverage the MDP framework to develop practical scheduling algorithms and optimize the receiver video quality for stored scalable video transmission over wireless channels. First, we propose a tractable MDP formulation based on a reasonable approximation of the state space. Near optimal scheduling policies can be derived from this MDP formulation. Second, we propose a scheduling algorithm that substantially simplifies the MDP-based scheduling policy as it requires only limited information regarding the channel state dynamics. Third, we prove an upper bound on the achievable video quality of all possible scheduling algorithms. Finally, we provide simulation results that show, under different channel conditions, the performance of the proposed scheduling algorithms is indeed very close to the upper bound.

Our contributions are summarized as follows:

\begin{enumerate}
  \item{\it An MDP formulation is proposed to facilitate the design of adaptive scheduling policies for stored video transmission.} \textcolor{black}{In this paper, we focus on stored video transport, where video sequences have been encoded and stored on a video server before transmission. This is quite different from real-time video transmission where video frames are generated in real-time. The video scheduler can select any data from the video sequence and send the data to the receiver buffer. Thus, the number of possible receiver buffer states can be effectively regarded as infinite.} Because the performance of the scheduling policy depends on the receiver buffer state, the policy needs to be optimized over an infinitely large state space and the scheduling problem is intractable. In this paper, by applying reasonable restrictions on the set of scheduling policies considered in our MDP formulation, we prove that optimizing the transmission policy is equivalent to solving a semi-Markov decision problem on a finite state set (see Section. \ref{sec:formulation}). Based on this result, near-optimal scheduling policies can be derived using the proposed MDP formulation.
  \item{\it A near-optimal and on-line scheduling algorithm is proposed.} In most cases, models for channel dynamics are not available. By simplifying the channel model and the scheduling decision of the MDP formulation, we devise an on-line scheduling algorithm which, unlike the MDP-based policy, only requires limited measurable knowledge of the channel dynamics. Simulation results show that the proposed on-line algorithm performs nearly as well as the MDP-based scheduling policy.
  \item{\it Performance optimality is justified.} To assess the performance of the proposed scheduling algorithms, an upper bound on the achievable video quality for adaptive scheduling is proved. Simulation results show that both the MDP-based scheduling policy and the proposed on-line scheduling policy perform close to the upper bound.
\end{enumerate}

\subsection{Related Work}
Adaptive video data scheduling is an important topic of research \cite{SoftARQ,Cuetos03, ChouMiao06,ChakChou02,ChanguelMM10,FuSchaar09,Mehaela10,Fu10,Structural12,ChaHeaBovdeV11}. In \cite{SoftARQ}, adaptive video transmission over a packet erasure channel was studied by modeling the buffer state as a controlled Markov chain. \textcolor{black}{In \cite{Cuetos03}, an average-rate constrained MDP formulation was proposed to optimize the quality of error concealed videos at the receiver. For time-varying wireless channels, the amount of data that can be scheduled during a time slot is limited by the channel capacity at the slot. Only considering the constraint of the average transmission rate is insufficient.} In \cite{ChouMiao06}, an MDP-based scheduling algorithm was proposed for video transmission over packet loss networks. This work was further extended for wireless video streaming in \cite{ChakChou02}, where the wireless channel was modeled as a binary symmetric channel. This channel model can only be justified for fast fading channels, where the coherence time is much less than the delay constraint. In that case, interleaving can be applied without violating the delay constraint, and the channel will appear as an i.i.d channel. For slow fading channels such as those considered here, the bit error rate cannot be modeled as a constant. \textcolor{black}{In \cite{ChanguelMM10}, adaptive scheduling of scalable videos was studied using an MDP model. The reward of each frame slot was defined as a utility function of the buffer state and the transmission rate. A foresighted scheduling policy was derived to maximize the long-term reward over all frame slots. Comparing with a scheduling method that myopically maximizes the reward of each individual frame slot, the proposed scheduling algorithm improved the video quality significantly.} In \cite{FuSchaar09},\cite{Mehaela10}, \cite{Fu10} and \cite{Structural12}, reinforcement learning frameworks were proposed for wireless video transmission. Their proposed algorithms were based on MDP using a discounted-reward maximization formulation. The transmitter learns the characteristics of the channel and the video sequence during the transmission process. The scheduling policy is updated according to the learned characteristics. In our previous work \cite{ChaHeaBovdeV11}, an infinite-horizon average-reward maximization MDP formulation was proposed. The channel characteristics, unlike in this paper, were assumed to be known.

The most closely related prior work is \cite{ChakChou02,ChanguelMM10,FuSchaar09,Mehaela10,Fu10} and \cite{Structural12} which focus on scalable video transmission over wireless channels. Our work contrasts with these as follows:
\begin{itemize}
  \item {\it An Infinite-State Space Problem for Stored Video Streaming.} \textcolor{black}{For real-time video transmission, the number of video frames that are ready for transmission is finite because later frames have not yet been generated at the video source. Therefore, the scheduler only needs to select data from a finite set of frames \cite{ChakChou02,ChanguelMM10,FuSchaar09,Mehaela10,Fu10}. In this paper, we focus on stored video streaming, where all the video frames have been encoded before transmission. In this case, the scheduler is allowed to select data from any video frame and the number of possible receiver buffer states is therefore infinitely large. In this paper, we construct a finite-state MDP model and show that the optimal policy derived from this MDP model is also optimal for the original infinite-state problem.}
  \item  {\it Channel Model.} We focus on slow-fading wireless channels experienced by pedestrian users. In the channel model of \cite{ChakChou02}, the bit error probability of the channel was assumed constant. This assumption can only be justified for fast fading channels, where the channel coherence time is much less than the delay constraint in video applications. In that case, interleaving can be applied without violating the delay constraint, and the channel will appear to have i.i.d. bit errors. For slow fading channels, where the coherence time is much longer, it is impossible to apply interleaving over many coherence periods due to the delay constraint. In this case, i.i.d. models are no longer suitable because they do not capture information regarding channel variations. In contrast, the algorithm proposed in this paper explicitly considers channel state variation in scheduling.
  \item {\it Optimization objective.} Most of the existing MDP-based scheduling algorithms are based on a {\it utility} function as the optimization objective \cite{ChanguelMM10,FuSchaar09,Mehaela10,Fu10}. \textcolor{black}{The utility function is usually written as a weighted sum of the transmission bit rate and the amount of buffered data. The weights assigned to each component of the summation, to some extent, reflect their importance, but are heuristically chosen.} The resulting utility function cannot accurately indicate the quality of played out frames. Here, instead of optimizing a utility function, we directly optimize the quality of the video frame played out in each frame slot.
  \item {\it Non-availability of channel state dynamics.} In a practical wireless video transmission application, models for the dynamics of the channel state are typically unavailable. To address this problem, a reinforcement learning algorithm can be employed to learn a good policy from making wrong scheduling actions \cite{FuSchaar09,Mehaela10,Fu10}. Video quality, however, will be degraded during the learning period, which can be as long as tens of seconds. We propose an adaptive alternative to such reinforcement learning methods, that only uses the channel coherence time and current channel throughput which are easy to measure in practice. The performance of the proposed algorithm is very close to a derived performance upper bound.
\end{itemize}
\subsection{Organization of Paper}

This paper is organized as follows. The system model is introduced in Section \ref{sec:sysmodel}. The assumptions we make about the video codec and the rate-distortion model are described in Section \ref{sec:sysmodel}. In Section \ref{sec:formulation}, the MDP formulation and the performance upper bound are proposed. A near-optimal on-line scheduling algorithm is introduced and validated by simulations in Section \ref{sec:nearoptimal}. Section \ref{sec:conclusion} concludes the paper.

\section{System Model}
\label{sec:sysmodel}
In this section, we describe the wireless video system to be considered. Then, we present our video codec configuration and introduce the rate-distortion model.

We briefly introduce some notation used in the paper. $\bf A$ and $\bf a$ are examples of a matrix and a vector, respectively. $\mathcal A$ is a set. $|\mathcal A|$ is the cardinality of set $\mathcal A$. $\lceil\cdot\rceil$ is the ceiling function. $\mathbb P(\cdot)$ is the probability measure and $\mathbb E[\cdot]$ is the expectation. $\mathbb N=\{0,1,2,\cdots\}$ is the set of non-negative integers. Other frequently used notation is summarized in Table \ref{tab:notations}.

\begin{table}[!t]
\renewcommand{\arraystretch}{1.3}
\caption{FREQUENTLY USED NOTATION.}
\label{tab:notations}
\centering
\begin{tabular}{|l|p{6.7cm}|}
%\hline
%{\bf Notations}&{\bf Descriptions}\\
%\hline
\hline
$F^\mathrm{intra}$&Number of frames in a intra period.\\

$F^\mathrm{GOP}$&Number of frames in a GOP.\\

$L$&Number of MGS layers.\\

$z_t$& The amount of received data for the frame played out at $t$.\\

$\omega^k_\ell$& The amount of data in the $\ell^\mathrm{th}$ layer of a type-$k$ frame.\\

$d_\ell$& The distortion when the $\ell^\mathrm{th}$ layer is correctly received.\\

$\mathrm{d}^k(z_t)$& The rate-distortion model for type-$k$ frames.\\

$\mathrm{\widehat d}^k(z_t)$& The concave envelopes of $\mathrm{d}^k(z_t)$.\\

$X_t$& The transmission bit rate at $t$.\\
$Y_t$& The packet error rate at $t$.\\
$R_t$& The channel throughput at $t$.\\

%$\hat{r_t}$& $\hat{r_t}=x_t(1-y_t)$ is the estimation of $r_t$.\\

$r^\mathrm{avg}$& The average channel throughput.\\

$C_t$&The channel state at $t$.\\
$V_t$&The buffer state at $t$\\
$S_t$& The system state at $t$.\\

%\hline
%$\bf P^c$& The transition probability matrix of the channel states.\\
%\hline
%$\mathbb P_{\mu}(\cdot|\cdot)$& System state transition probabilities under scheduling policy $\mu$.\\
\hline
\end{tabular}
\end{table}

\subsection{System Overview}
\label{sec:sysoverview}
\begin{figure}[ht]
\centering
\includegraphics[width=3in]{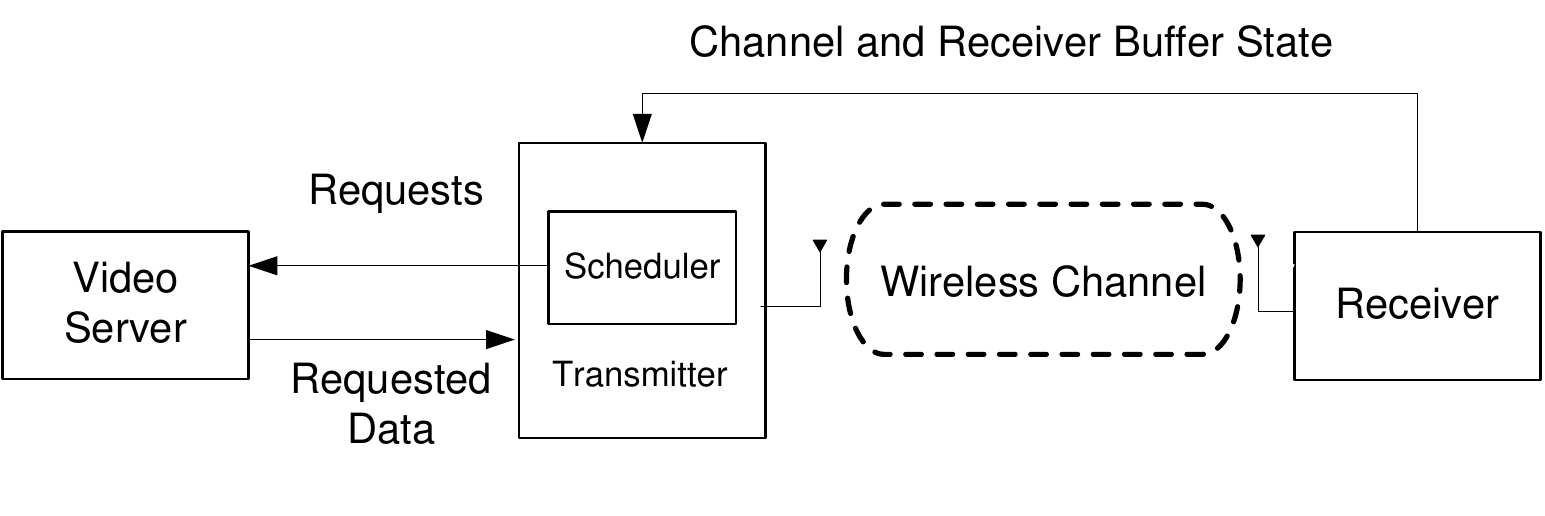}
\caption{Dynamic scheduling system for wireless video transmission.}
\label{fig:system}
\end{figure}

We consider a time-slotted system that transmits scalable videos over a slow fading wireless channel. The video sequence is encoded with a quality-scalable video encoder and is stored in a video server. The video server transmits video data to a mobile user via a wireless transmitter. The duration of each frame $\Delta T$ is called a frame slot. In each frame slot, the server sends some video data upon request of a scheduler at the wireless transmitter. This data is packetized at the wireless transmitter for physical layer transmission. The channel and receiver buffer state is sent to the scheduler via a feedback channel with negligible delay. The scheduler operates according to a policy that maps the channel and receiver buffer state to the scheduling action (see Fig. \ref{fig:system}).

\textcolor{black}{In wireless communication systems such as 3GPP, using the technique of limited feedback, channel state information measured at the receiver can be fed back to the transmitter via a control channel \cite{LF04}\cite{LF08}\cite{3gpp}. The delay of the feedback channel is typically much smaller than a frame slot. For example, the feedback delay in 3GPP is 6ms \cite{3gpp}, which is much shorter than the 33ms frame slot of 30fps videos. Similarly, the video packets received in each slot can also be acknowledged via a control channel with negligible delay. Therefore, similar to most of existing MDP formulations such as \cite{ChanguelMM10,FuSchaar09,Mehaela10,Fu10}, we assume the feedback is instantaneous. For the case where feedback delay is longer than a frame slot, please refer to \cite{Changuel12}.}

We assume that the link between the video server and the wireless transmitter is not the bottleneck for transmission to the mobile. Thus, from the perspective of the wireless transmitter, the whole video sequence is available for transmission. We also assume that the physical layer channel state information is available at the transmitter and that the modulation and coding scheme (MCS) is determined by a given physical layer link-adaptation policy.

\subsection{Video Codec Configuration}
\label{sec:codec}
\begin{figure}[ht]
\centering
\includegraphics[width=3.3in]{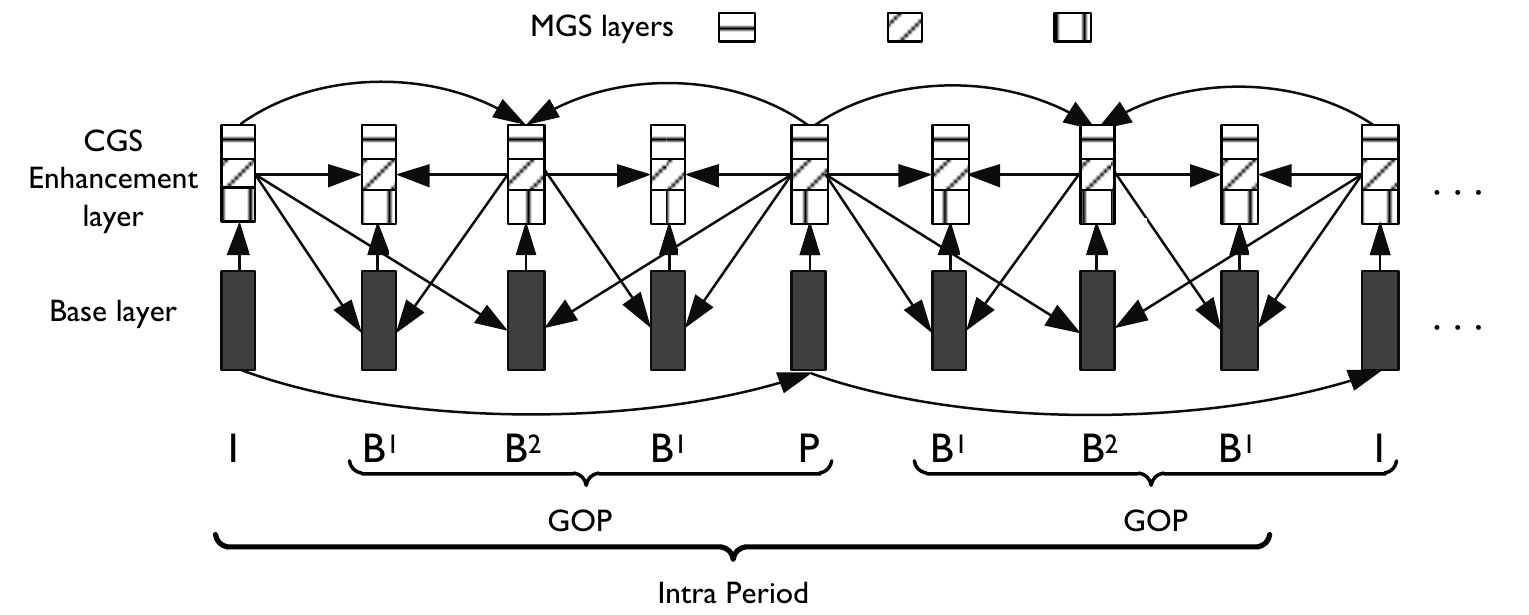}
\caption{An illustration of the encoder prediction structure considered in this paper. The prediction order is indicated by arrows. The length of intra period is $F^\mathrm{intra}=8$ and the GOP length is $F^\mathrm{GOP}=4$. The CGS enhancement layer is partitioned into 3 MGS layers.}
\label{fig:structure}
\end{figure}

We assume that the video sequence is encoded by an H.264/SVC video encoder. The video frames are uniformly partitioned into intra periods. Every intra period has $F^\mathrm{intra}$ frames and is further partitioned uniformly into group of pictures (GOP
s). Each GOP has $F^\mathrm{GOP}$ frames. They are encoded using the ``Hierarchical B" prediction structure \cite{SVC}, in which video frames are hierarchically organized into $T$ temporal layers as shown in Fig.~\ref{fig:structure}. The last frame in each GOP is called a {\it key picture}. These key pictures form the $0^\mathrm{th}$ temporal layer. There are two types of key pictures: $I$ frames and $P$ frames. The first picture in a intra period is called an $I$ frame, which is encoded without referring other frames. The other key pictures are $P$ frames. Each $P$ frame is encoded using a preceding key pictures as reference. All the frames in higher temporal layers are $B$ frames. A $B$ frame in the $\tau^\mathrm{th}$ temporal layer is encoded using the preceding frame and the succeeding frame in the lower temporal layers as reference. In the following, we call a frame in the $\tau^\mathrm{th}$ temporal layer a $B^\tau$ frame, where $\tau\geq1$ (see Fig.~\ref{fig:structure}).

Every frame is encoded into a base layer and a coarse grain scalability (CGS) layer. The base layer of an $I$ frame is encoded independently. The base layer of a $P$ frame is predictively encoded using the base layer of the preceding key picture. The CGS layer of all key picture are predictively encoded using their respective base layers. For a $B$ frame, its base layer is encoded using the CGS layers of its reference frames. Its CGS layer is encoded using both its base layer and the CGS layers of its reference frames (see Fig. \ref{fig:structure}).

The CGS layer of each frame is further partitioned into $L$ MGS layers. Each MGS layer contains a portion of the CGS layer data. Thus, the more MGS layers are received, the higher decoding quality can be achieved. In the following, we call the base layer and the MGS layers {\it quality layers}. We focus on adaptive scheduling of the quality layers in a video stream. The temporal scalability is not exploited.

In this paper, we only consider one CGS enhancement layer. In H.264/SVC, multiple CGS layer is supported and each CGS layer can be partitioned into several MGS layers. The switch between CGS layers, however, is only possible at instantaneous decoder refresh (IDR) frames, which are separated from each other by several intra periods. Therefor, CGS cannot support frame-by-frame rate adaptation. Since the coherence time of wireless channels is much shorter than a intra period, flexible rate adaptation can only be achieved by MGS, which allows to vary the number of quality layer for each frame. Here, we consider frame-by-frame adaptive scheduling of the MGS layers within a single CGS enhancement layer. For the video streams that contain multiple CGS enhancement layers, our scheduling algorithm can be applied to conduct adaptation in one of the CGS enhancement layers while treating all lower layers as the base layer. In the following, we call the MGS layers {\it enhancement layers}.

\begin{figure}[ht]
\centering
\includegraphics[width=3.3in]{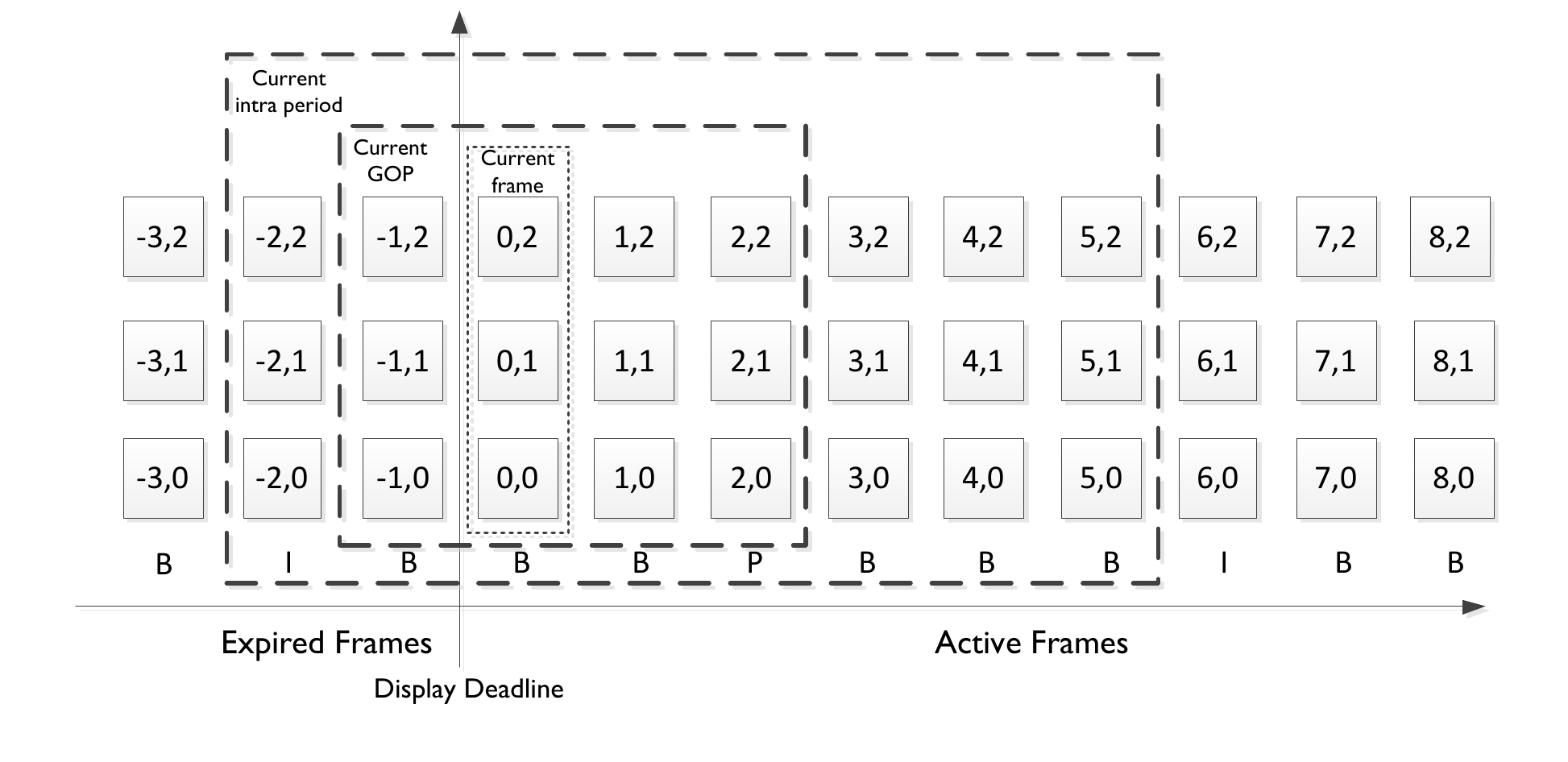}
\caption{Indices of data units when three quality layers are considered. At the beginning of each time slot, the frame with index $f=0$ is played out. All the data units in the figure shift left.}
\label{fig:index}
\end{figure}

Each frame has a playout deadline at the receiver. In the following, frames whose deadlines have expired are called expired frames, otherwise they are said to be active frames. The first active frame is called the ``current frame". The GOP that contains the current frame is called the ``current GOP". The intra period that contains the current frame is called the ``current intra period" (see Fig.~\ref{fig:index}). The frames in the current GOP are decoded together when the first frame of the GOP is displayed. At any point in time, frames are indexed relative to the current frame as shown in Fig. \ref{fig:index}. Each data unit is also tagged with a layer index $\ell$. The index for base layer is $\ell=0$ and the enhancement layers are index from 1 to $L$. The video data in the ${\ell}^\mathrm{th}$ layer of the $f^\mathrm{th}$ frame is called the $(f,\ell)^\mathrm{th}$ {\bf video data unit}.

\subsection{Rate-Distortion Model}
\label{sec:RDmodel}

\begin{figure}[ht]
\centering
\includegraphics[width=3in]{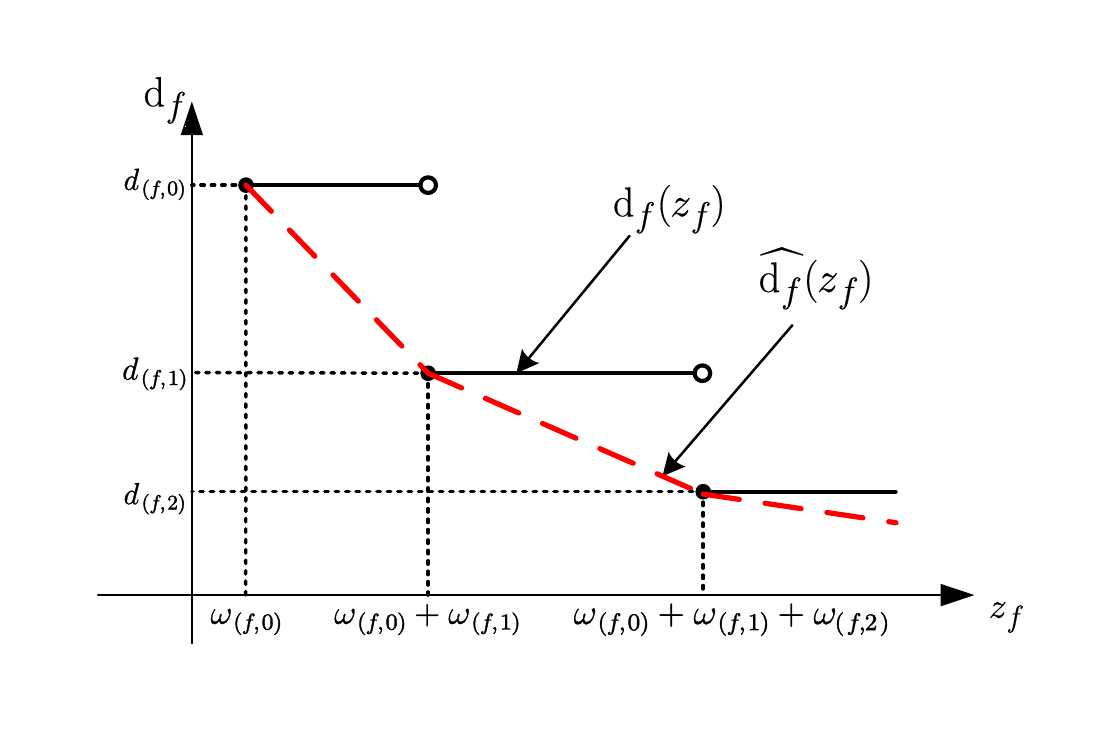}
\caption[Optional caption for list of figures]{
An illustration of the rate-distortion function $\mathrm{d}_f(z_f)$ for the $f^\mathrm{th}$ frame. The rate-distortion function $\mathrm{d}_f(z_f)$ is piecewise constant and right-continuous (solid). Its convex envelope $\mathrm{\widehat d}_f(z_f)$ is also shown (dashed).
}
\label{fig:RD}
\end{figure}

Let $z_f$ be the amount of received data for the $f^\mathrm{th}$ frame. The rate-distortion function $\mathrm{d}_{f}(z_{f})$ captures the quality of the frame when it is decoded, given all its predictors have been received. Let $\omega_{(f,\ell)}$ be the amount of data in the ${(f,\ell)^\mathrm{th}}$ data unit and $d_{(f,\ell)}$ be the distortion measured in mean square error (MSE) if the $0^\mathrm{th}\sim\ell^\mathrm{th}$ layers has been correctly received. As shown in Fig.~\ref{fig:RD}, since a data unit can be decoded only when all its associated data has been received, $\mathrm{d}_f(z_{f})$ is a piecewise constant and right-continuous function with jumps at $z_f=\sum_{\ell=0}^m{\omega_{(f,\ell)}}$, $m=0,1,\cdots,L$. Thus $d_{(f,\ell)}$ and $\omega_{(f,\ell)}$ characterize $\mathrm{d}_f(z_{f})$.

\textcolor{black}{In a real video sequence, for a given layer $\ell$, the rate-distortion characteristics $\omega_{(f,\ell)}$ and $d_{(f,\ell)}$ vary across frames. Let $\mathcal{K}=\{I,P,B^1,\cdots,B^{T}\}$ be the set of frame types. We model $\omega_{(f,\ell)}$ of type $k$ frames as i.i.d. realizations of a random variable $\Omega^k_\ell$, where $k\in\mathcal{K}$. Then, we use $\omega^k_\ell=\mathbb{E}[\Omega^k_\ell]$ as an estimate of $\omega_{(f,\ell)}$. Similarly, for a given layer $\ell$, we model $d_{(f,\ell)}$ as i.i.d. realizations of a random variable $D_\ell$. We use $d_\ell=\mathbb{E}[D_\ell]$ as an approximation of $d_{(f,\ell)}$. Here, we choose not to distinguish the frame types when modeling $d_{(f,\ell)}$. In a typical H.264/SVC video stream, the quantization parameters of the encoder are usually configured to minimize visually annoying quality fluctuations across different types of frames. Hence, for simplicity, we use a single random variable $D_\ell$ to model $d_{(f,\ell)}$ for all types of frames.}

Our rate-quality models $\mathrm{d}^k(z_f)$ for type-$k$ frames are constructed as piecewise constant functions with jumps at $z_f=\sum_{\ell=0}^{m}{\omega^k_\ell}$, $m=0,1,\cdots,L$. \textcolor{black}{For stored video transmission, which is the focus of this paper, since the transmitted video has already been encoded, the size of each data unit is thus available. The parameters $\{\omega^k_\ell, k\in\mathcal{K}\}$ can thus be estimated by averaging across frames. If the distortion characteristic $d_{(f,\ell)}$ is calculated when the video is encoded, the parameter $d_\ell$ can also be estimated by averaging $d_{(f,\ell)}$ across frames. If $d_{(f,\ell)}$ is not available, $d_\ell$ needs to be estimated on-line. For example, the quality of frames that have been decoded at the receiver can be fed back to the transmitter for estimation.}

\subsection{Streaming Setup}
\label{sec:channelModel}
We focus on scheduling for a slow fading channel. By slow fading, we mean that the coherence time of the channel is less than the duration of a intra period and larger than a frame slot. Assuming the mobile users are moving in a 1.5m/s walking speed and the carrier frequency is 2GHz, the Doppler spread is about 10Hz. The coherence time is about 100ms. A typical intra period duration is about 1 second and a frame slot is about 30ms. Hence, for pedestrian video users, wireless channels are slow fading.

As the channel state is stable during each frame slot, the scheduling decision is made on a frame-by-frame basis. At the beginning of each frame slot, a frame is played out, and video data units are scheduled for transmission. The scheduling action is defined as a set of ordered video data units
\textcolor{black}{
\begin{equation}
\label{eq:action}
 {\mathcal U}= \left\{({f_1},{\ell_1}),({f_2},{\ell_2}), \cdots ,({f_{|{\mathcal U}|}},{\ell_{|{\mathcal U}|}})\right\}.
\end{equation}}
When scheduling action $\mathcal U$ is taken, the associated data units are transmitted sequentially. Each scheduled data unit is packetized into physical layer packets and each packet is repeatedly transmitted, i.e., if packet error occur, until acknowledged.

\textcolor{black}{In this paper, we consider data unit level scheduling. If a packet level rate-quality model such as \cite{Fu10} is available, our MDP formulation can also be used to optimize the packet level scheduling policy.}

\section{Markov Decision Process-Based Model}
\label{sec:formulation}
In this section, we propose an MDP-based model to determine the near-optimal scheduling policy. To that end, we describe the scheduler's state space and the policies to be considered. We then show how to reduce the scheduling problem to a finite-state Markov decision problem using reasonable approximations. With the MDP-based model, the optimal scheduling policy is computed off-line via value iteration. The computed policy can then be used for on-line adaptive scheduling. To validate the optimality of the MDP-based scheduling policies, we develop a performance upper bound at the end of this section.

\subsection{Scheduling Policy and State Space}
\label{sec:simplestate}

Considering all possible scheduling actions makes defining the scheduling policy and representing the buffer state unmanageably complex. On one hand, to capture the buffer state, the frame index and the layer index of each received data unit need to be recorded. If we assume an infinite playback buffer, the number of received data units is not bounded. So we cannot represent all possible buffer states using a finite-dimensional space. On the other hand, not all possible scheduling policies need to be considered. For example, a quality layer of a frame should not be transmitted before the lower quality layers of the frame because a SVC decoder cannot decode a quality layer without the lower layers \cite{SVC}. Thus we need only consider those scheduling strategies that are not dominated and have potential to achieve good performance.

\textcolor{black}{Specifically, we consider scheduling policies under the following assumptions}:
\newtheorem{assumption}{\bf Assumption}
\begin{assumption}
\label{cond:precedent}
The scheduler always schedules the base layer data unit of a frame for transmission after the base layer data unit(s) of the reference frame(s). The scheduler always schedules the enhancement layer data unit of a frame for transmission after the data units of the lower layers.
\end{assumption}

\begin{assumption}
\label{cond:last}
\textcolor{black}{The scheduler always schedule enough amount of data such that the transmitter is kept transmitting during the whole slot.}
\end{assumption}

\begin{figure}[ht]
\centering
\subfigure[]{
\includegraphics[width=3.2in]{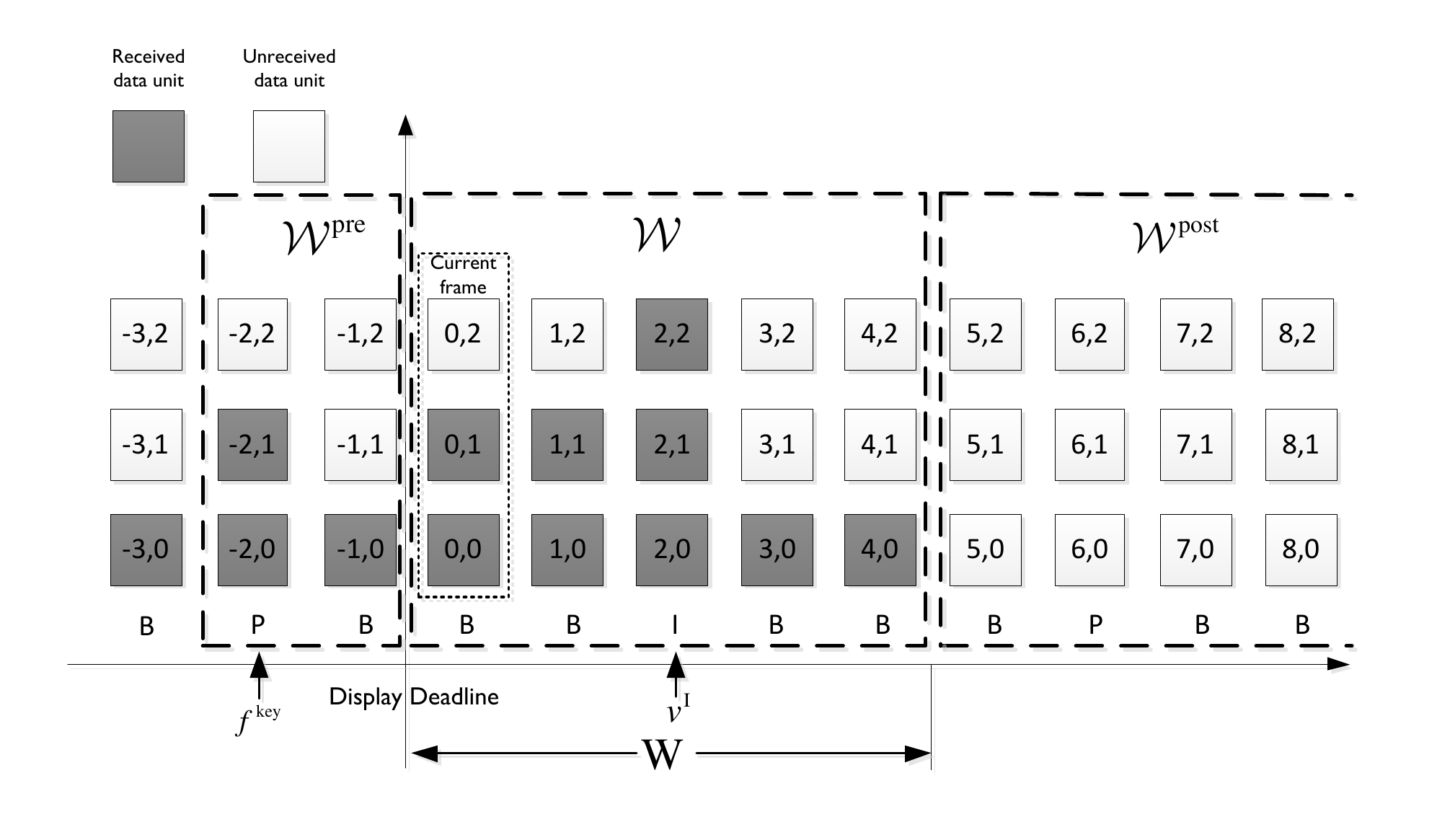}
\label{fig:bufferstate_noitr}
}
\subfigure[]{
\includegraphics[width=3.2in]{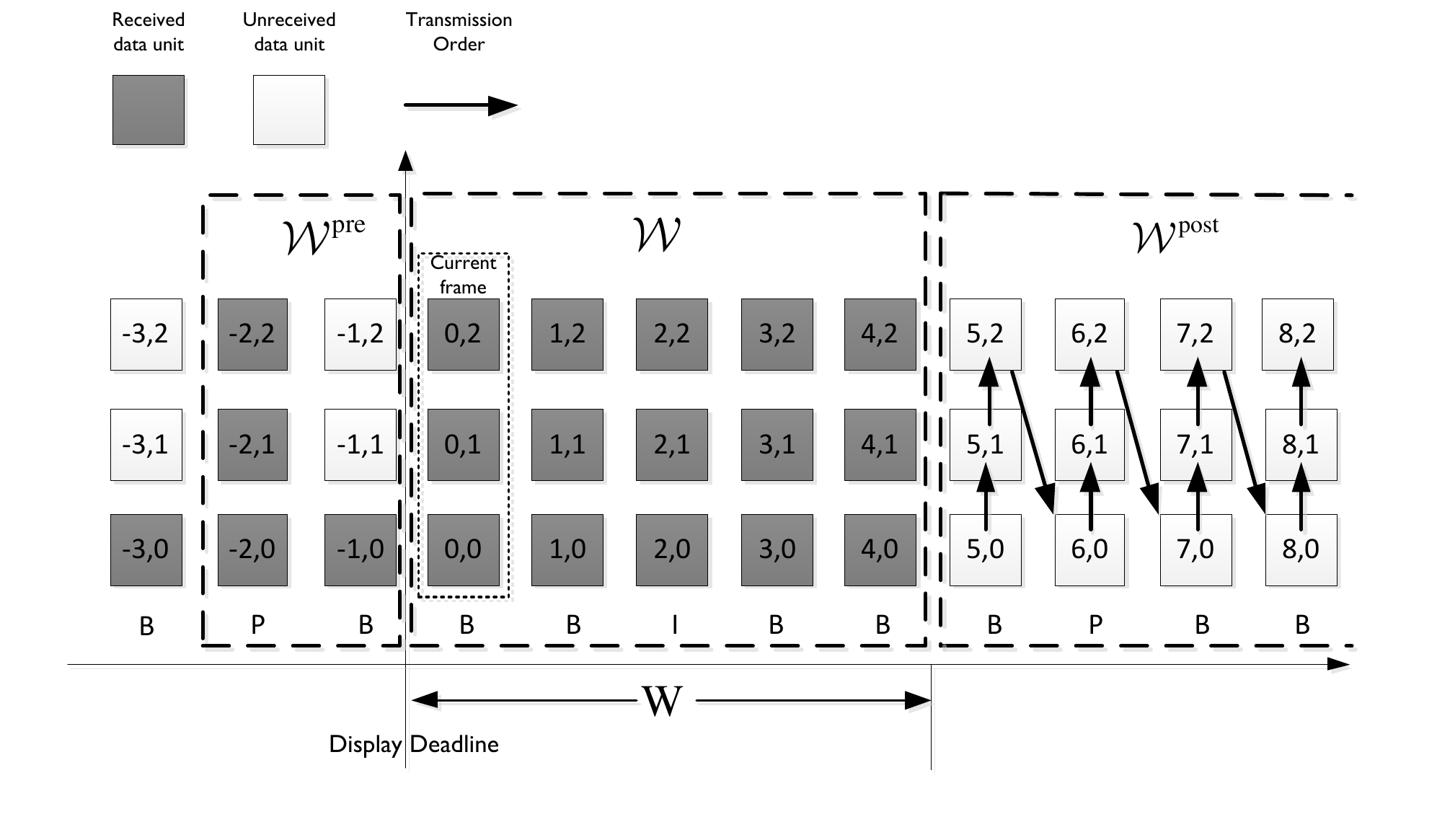}
\label{fig:neworder}
}
\caption[Optional caption for list of figures]{
\subref{fig:bufferstate_noitr}: An illustration of the receiver buffer state when $F^\mathrm{intra}=8$, $F^\mathrm{GOP}=4$, $L=2$, and $W=5$.  $v^\mathrm{I}=2$, ${\mathbf v}^\mathrm{pre}=(2,1)$, ${\mathbf v}^\mathrm{W}=(2,2,3,1,1)$, ${\mathbf v}^\mathrm{post}=(0,0,0)$. Note that because some data units in $\mathcal W$ have not been received, the data units in $\mathcal{W}^\mathrm{post}$ are not sent. \subref{fig:neworder}: The transmission order when the data in $\mathcal W$ and the associated predictors in ${\mathcal W}^\mathrm{pre}$ has been received.
\label{fig:assumption3}
}

\end{figure}

\begin{assumption}
\label{cond:window}\textcolor{black}{
We define three sets of data units: ${\mathcal W}^\mathrm{pre}$, $\mathcal W$ and ${\mathcal W}^\mathrm{post}$.
When the current frame is a $B$ frame, the set ${\mathcal W}^\mathrm{pre}$ contains the data units with frame index $f\in[f^\mathrm{key},-1]$, where $f^\mathrm{key}$ is the frame index of the last expired key picture (see Fig.~\ref{fig:bufferstate_noitr}). When the current frame is a key picture, we define ${\mathcal W}^\mathrm{pre}=\emptyset$. Note that ${\mathcal W}^\mathrm{pre}$ contains all the expired data units that are used to predict the frames in the current GOP. The set $\mathcal W$ contains the data units in all quality layers of the first $W$ frames, where $W$ is larger than $F^\mathrm{GOP}$. The set ${\mathcal W}^\mathrm{post}$ contains the remaining active data units. We assume the scheduler first sends the data units in $\mathcal W$. Then, if all the data in $\mathcal W$ and the predictors in ${\mathcal W}^\mathrm{pre}$ have been received, the policy greedily schedules all $1+L$ quality layers of the frames in ${\mathcal W}^\mathrm{post}$, i.e., starts transmitting the next frame in ${\mathcal W}^\mathrm{post}$ only when all the layers of the preceding frame have been received (see Fig.~\ref{fig:neworder}).}
\end{assumption}

\begin{assumption}
\label{cond:decoding}\textcolor{black}{
In each slot, the scheduler only schedules data for the frames that have not been decoded.
}
\end{assumption}
Assumption \ref{cond:precedent} ensures that the transmission order is compatible with the prediction order given in Section \ref{sec:codec}. \textcolor{black}{Assumption \ref{cond:last} ensures the transmitter will not be idle during a slot and the capacity of the channel is fully exploited.} Assumption \ref{cond:window} stems from the fact that, when many frames are buffered at the receiver, the scheduler can transmit more enhancement layers because there is sufficient time before the frames are played out. In other words, if all quality layers of $W$ frames have been received, there is not need to worry about the channel capacity variation in the future. As will be discussed in Section.~\ref{sec:finiteformulation}, this assumption helps to simplify the policy optimization problem. It should be noted that policies under Assumption \ref{cond:window} are different from the sliding window policies defined in \cite{ChouMiao06}. Indeed, our scheduling policy allows the transmitter to transmit data units outside the window. Assumption \ref{cond:decoding} ensures that the transmitter does not waste resources on the frames which have been decoded.

\begin{remark}
\textcolor{black}{The window size $W$ provides a tradeoff between complexity and optimality. The larger the window, the less constrained the control policy but the higher complexity\footnote{In our simulations, we find that setting $W=9$ is sufficient.}. We note that, although the frames in the current GOP are played out sequentially, they are decoded together. According to Assumption \ref{cond:decoding}, if $W\geq F^\mathrm{GOP}$, the frames in $\mathcal W$ have all been decoded and the scheduler cannot schedule any data from $\mathcal W$. Therefore, we set $W\geq F^\mathrm{GOP}$.}
\end{remark}

We define the overall buffer state space $\mathcal V$ via four sets $\mathcal V^\mathrm{I}$, $\mathcal {V^{\mathrm{pre}}}$, $\mathcal {V^{\mathrm{W}}}$, and  $\mathcal{V}^{\mathrm{post}}$, where $\mathcal V=\mathcal {V^\mathrm{I}}\times\mathcal V^\mathrm{pre}\times\mathcal{V}^{\mathrm{W}}\times\mathcal{V}^{\mathrm{post}}$. The set $\mathcal V^\mathrm{I}$ records the types and playout deadlines of the frames in the buffer. The sets $\mathcal {V^{\mathrm{pre}}}$, $\mathcal {V^{\mathrm{W}}}$, and  $\mathcal{V}^{\mathrm{post}}$ describe the states of the frames in ${\mathcal W}^\mathrm{pre}$, $\mathcal W$ and ${\mathcal W}^\mathrm{post}$, respectively.
\begin{description}
  \item[$\mathcal V^\mathrm{I}$:] \textcolor{black}{We define $v^\mathrm{I}$ as the frame index of the active $I$ frame with the earliest playout deadline. Since the prediction structure is assumed to be the same for all intra periods, $v^\mathrm{I}$ determines the types and playout deadlines of all the frames in the receiver buffer.}
  \item[$\mathcal {V^{\mathrm{pre}}}$:] If the current frame is a $B$ frame, the state space $\mathcal {V^{\mathrm{pre}}}$ is defined as a vector ${\mathbf v}^{\mathrm{pre}}=(b^\mathrm{pre}_{f^\mathrm{key}}, \cdots, b^\mathrm{pre}_{-1})$, where $b^\mathrm{pre}_f$ is the number of the received quality layers in the $f^\mathrm{th}$ frame and $f^\mathrm{key}$ is the frame index of the last expired key picture. If the current frame is a key picture, $\mathcal{W}^\mathrm{pre}=\emptyset$ and we define ${\mathbf v}^{\mathrm{pre}}=-1$.
  \item[$\mathcal {V^{\mathrm{W}}}$:] Similar to $\mathcal {V^{\mathrm{pre}}}$, we define the buffer state space for $\mathcal{W}$ as a vector ${\mathbf v}^{\mathrm{W}}=(b^\mathrm{W}_0, \cdots, b^\mathrm{W}_{W-1})$, where $b^\mathrm{W}_f$ is the number of the received quality layers in the $f^\mathrm{th}$ frame.
  \item[$\mathcal{V}^{\mathrm{post}}$:] The set $\mathcal{W}^\mathrm{post}$ contains infinite number of frames. Therefore, recording the number of data units received for each frame is impossible. We note that, when Assumption \ref{cond:window} is enforced, the number of data units received in $\mathcal{W}^\mathrm{post}$ must be non-increasing in the frame index. Hence, we only need record the total number of received data units for each layer. We define the buffer state space of $\mathcal{W}^\mathrm{post}$ as a $1+L$-dimensional vector ${\mathbf v}^{\mathrm{post}}=(b^\mathrm{post}_0, b^\mathrm{post}_1, \cdots, b^\mathrm{post}_L)$, where $b^\mathrm{post}_\ell$ is the number of the received data units in the $\ell^\mathrm{th}$ layer of $\mathcal{W}^\mathrm{post}$. Because the receiver buffer size is assumed to be large, i.e., essentially infinite, $b^\mathrm{post}_\ell$ is unbounded. Thus $\mathcal{V}^\mathrm{post}=\mathbb{N}^{1+L}$, where $\mathbb{N}=\{0,1,\cdots,\infty\}$.
\end{description}
With the above definition, buffer state $\mathbf{v}=(v^\mathrm{I}, \mathbf{v}^\mathrm{pre}, \mathbf{v}^\mathrm{W}, \mathbf{v}^\mathrm{post})$ contains all the information that is relevant to the quality of frames in the receiver buffer.

In \cite{ZhangKassam99} and \cite{WangChang96}, it is shown that a first-order finite state Markov chain (FSMC) can be used to describe the first-order channel state transition probabilities for Rayleigh fading channels. First-order FSMC models have also been validated in \cite{LinTseng05} and \cite{SuLingVo98} by wireless channel measurements in urban areas. In our MDP-based model, we employ a first-order FSMC to describe the dynamics of the channel state.

We denote by $x$ the transmission rate of the transmitter, i.e., the number of bits transmitted in a time slot $\Delta T$. We denote by $y$ the packet error rate of the channel. We define the channel state as ${\mathbf c}=(x,y)$. The channel state space is $\mathcal {C}=\left\{{\mathbf c}^1,..., {\mathbf c}^{|\mathcal C|}\right\}$, where ${\mathbf c}^i=(x^i,y^i)$ is the $i^\mathrm{th}$ channel state. The state transition matrix ${\bf P}^\mathrm{c}$ is a $|\mathcal C|\times|\mathcal C|$ matrix with entry ${\mathbf P}^\mathrm{c}_{i,j}=\mathbb P({\mathbf c}^j|{\mathbf c}^i)$ being the transition probability from state ${\mathbf c}^i$ to ${\mathbf c}^j$.

The system state space $\mathcal S$ is defined as the product of the buffer state space $\mathcal V$ and the channel state space $\mathcal C$. For each state ${\mathbf s}\in \mathcal S$, we define a feasible control set $\mathcal U_{{\mathbf s}}$ that contains all the scheduling actions (see Equation \eqref{eq:action}) complying with all the three assumptions. The state ${\mathbf s}$ contains all the information about the receiver buffer and the channel. The transmitter must decide which action in $\mathcal U_{{\mathbf s}}$ to take in order to minimize the distortion. We define the scheduling policy $\mu(\cdot)$ as the mapping from the system state ${\mathbf s}$ to an action in $\mathcal U_{{\mathbf s}}$. Under given scheduling policy $\mu$, the system state transit as a controlled Markov chain. The state transition probability ${\mathbb P}_\mu(\cdot|\cdot)$ is determined by the scheduling policy $\mu$ (see Appendix \ref{sec:transition} for detail). In the following sections, we show how to optimize the scheduling policy $\mu(\cdot)$.

\subsection{Optimization Objective}
Since the channel condition is modeled as a random process, we denote by $\left(C_t, V_t, S_t\right)_{t\in\mathbb{N}}$ the random processes modeling channel, buffer and system state, respectively. Accordingly, we denote $S=\lim_{t \to +\infty}S_t$. \textcolor{black}{We define a function $\mathrm{d}(\mathbf{s})$ of state $\mathbf{s}$ as the estimated distortion of the frame that is played out at state $\mathbf{s}$. \footnote{Since in each slot, a frame is played out before the scheduling actions are taken. Therefore, $\mathrm{d}(\mathbf{s})$ is not a function of the actions taken at the state $\mathbf{s}$.} Our aim is to find an optimal policy $\mu^*(\cdot)$ that minimizes the expectation of distortion, i.e.,
\begin{equation}
J_{\mu}={\mathbb E}_{\mu}\left[\mathrm{d}(S)\right],	
\end{equation}
where $\mathbb{E}_\mu[\cdot]$ is the expectation over the stationary distribution of the controlled Markov chain under policy $\mu$. }

\textcolor{black}{We now introduce the definition of $\mathrm{d}(\mathbf{s})$. If the displayed frame is a key picture ($I$ frame or $P$ frame), we estimate its distortion using the rate-distortion model in Section~\ref{sec:RDmodel} as
\begin{equation}
\label{eq:key_quality}
\mathrm{d}(\mathbf{s})=\left\{
\begin{array}{lcl}
\mathrm{d^I}\left(\mathrm{z}(\mathbf{s})\right)&: &\text{for I frames}\\
\mathrm{d^P}\left(\mathrm{z}(\mathbf{s})\right)& :&\text{for P frames},	
\end{array}
\right.
\end{equation}
where $\mathrm{z}(\mathbf{s})$ denotes the amount of received data for the displayed frame at state $\mathbf{s}$. If the displayed frame is a $B$ frame, which is encoded using all the $1+L$ layers of its reference frames as predictor, the distortion cannot be directly estimated using the rate-distortion model in Section~\ref{sec:RDmodel} because the loss in the enhancement layers of its reference frames causes encoder-decoder predictor mismatch, which is also known as drift in SVC \cite{SVC}. We employ the model proposed in \cite{GaoWen} to take into account the distortion due to drift. Let  $\left\{\widehat{v_i^\mathrm{ref}},~i\in{1,2}\right\}$ denote the predictor for the $B$ frame at the encoder, i.e., the pixel value of the $i^\mathrm{th}$ reference frame with all the $1+L$ layers. Let $\left\{\widetilde{v_i^\mathrm{ref}},~i\in{1,2}\right\}$ be the the predictor for the $B$ frame at the decoder, i.e., the pixel value of the $i^\mathrm{th}$ reference frame with all the received quality layers. The drift of the reference frame is thus $\epsilon_i^\mathrm{dft}=\widetilde{v_i^\mathrm{ref}}-\widehat{v_i^\mathrm{ref}}$. In \cite{GaoWen}, it is shown that the MSE of a type-$B^\tau$ frames can be estimated as:
\begin{equation}
\label{eq:gaowen}
\tilde{\mathrm{d}}(\mathbf{s})=\mathrm{d^{B^\tau}}\left(\mathrm{z}(\mathbf{s})\right)+\frac{1}{4}\mathbb{E}\left[(\epsilon_1^\mathrm{dft})^2\right]+\frac{1}{4}\mathbb{E}\left[(\epsilon_2^\mathrm{dft})^2\right]+\frac{1}{2}\mathbb{E}\left[\epsilon_1^\mathrm{dft}\epsilon_2^\mathrm{dft}\right],
\end{equation}
where $\mathrm{d^{B^\tau}}\left(\mathrm{z}(\mathbf{s})\right)$ is the rate-distortion function defined in Section~\ref{sec:RDmodel} and the other terms on the right hand side are the distortions due to drift. Since $\mathbb{E}[(\epsilon_1^\mathrm{dft}-\epsilon_2^\mathrm{dft})^2]=\mathbb{E}[(\epsilon_1^\mathrm{dft})^2]+\mathbb{E}[(\epsilon_1^\mathrm{dft})^2]-2\mathbb{E}[\epsilon_1^\mathrm{dft}\epsilon_2^\mathrm{dft}]\geq0$, we have $\mathbb{E}[\epsilon_1^\mathrm{dft}\epsilon_2^\mathrm{dft}]\leq\frac{1}{2}\mathbb{E}[(\epsilon_1^\mathrm{dft})^2]+\frac{1}{2}\mathbb{E}[(\epsilon_2^\mathrm{dft})^2]$.
Thus, $\tilde{\mathrm{d}}(\mathbf{s})$ is upper bounded by $\mathrm{d^{B^\tau}}(\mathrm{z}(\mathbf{s}))+\frac{1}{2}\mathbb{E}[(\epsilon_1^\mathrm{dft})^2]+\frac{1}{2}\mathbb{E}[(\epsilon_2^\mathrm{dft})^2]$.
We use this upper bound as a proxy of the $B$ frame's distortion in our MDP model. The function $\mathrm{d}(\mathbf{s})$ is defined as
\begin{equation}
\label{eq:B_MSE}
\mathrm{d}(\mathbf{s})=\mathrm{d^{B^\tau}}\left(\mathrm{z}(\mathbf{s})\right)+\frac{1}{2}\mathbb{E}\left[(\epsilon_1^\mathrm{dft})^2\right]+\frac{1}{2}\mathbb{E}\left[(\epsilon_2^\mathrm{dft})^2\right].
\end{equation}
The term $\mathbb{E}[(\epsilon_1^\mathrm{dft})^i]$ is estimate from the distortion of the reference frame as follows. Let $v_i^\mathrm{ref}$ be the original pixel value of the reference frame before encoding. The decoding error of the reference frame is thus $\epsilon^\mathrm{ref}_i=\widetilde{v_i^\mathrm{ref}}-v_i^\mathrm{ref}=(\widetilde{v_i^\mathrm{ref}}-\widehat{v_i^\mathrm{ref}})+(\widehat{v_i^\mathrm{ref}}-v_i^\mathrm{ref})$, where $\widetilde{v_i^\mathrm{ref}}-\widehat{v_i^\mathrm{ref}}=\epsilon^\mathrm{dft}_i$ is the distortion due to drift and $\widehat{v^\mathrm{ref}}-v_i^\mathrm{ref}$ is the distortion due to encoding. Assuming $\widetilde{v_i^\mathrm{ref}}-\widehat{v_i^\mathrm{ref}}$ and $\widehat{v_i^\mathrm{ref}}-v_i^\mathrm{ref}$ are uncorrelated\footnote{This assumption is empirically true. We calculated the correlation coefficient of $\widetilde{v_i^\mathrm{ref}}-\widehat{v_i^\mathrm{ref}}$ and $\widehat{v_i^\mathrm{ref}}-v_i^\mathrm{ref}$ using the frames of test sequence ``foreman", ``Paris", and ``bus". The average correlation coefficient is 0.05.}, we have $\mathbb{E}[(\epsilon_i^\mathrm{ref})^2]=\mathbb{E}[(\epsilon_i^\mathrm{dft})^2]+\mathbb{E}[(\widehat{v_i^\mathrm{ref}}-v_i^\mathrm{ref})^2]$.
Since the $B$ frame is predicted by the $L^\mathrm{th}$ enhancement layer of the reference frame, we have  $\mathbb{E}[(\widehat{v_i^\mathrm{ref}}-v_i^\mathrm{ref})^2]=d_L$. Denoting by $\mathrm{d}_i^\mathrm{ref}(\mathbf{s})=\mathbb{E}[(\epsilon_i^\mathrm{ref})^2]$ the distortion of the reference frame, we have
\begin{equation}
\label{eq:B_dft}
\mathbb{E}[(\epsilon_i^\mathrm{dft})^2]=\mathrm{d}_i^\mathrm{ref}(\mathbf{s})-d_L.
\end{equation}
Substituting \eqref{eq:B_dft} into \eqref{eq:B_MSE}, we have
\begin{equation}
\label{eq:B_quality}
\mathrm{d}(\mathbf{s})=\mathrm{d^{B^\tau}}\left(\mathrm{z}(\mathbf{s})\right)+\frac{1}{2}\left[\mathrm{d_1^{ref}}(\mathbf{s})+\mathrm{d_2^{ref}}(\mathbf{s})\right]-d_L.
\end{equation}
The distortion of the reference frame $\mathrm{d}_i^\mathrm{ref}(\mathbf{s})$ can be recursively estimated using \eqref{eq:key_quality} and \eqref{eq:B_quality}. Because the prediction structure is acyclic, the recursion terminates when the reference frame is a key picture and \eqref{eq:key_quality} applies.}

\textcolor{black}{It should be noted that, in \eqref{eq:B_quality}, the distortion is overestimated using an upper bound of \eqref{eq:gaowen}. As will be shown by the simulation results in Section.\ref{sec:mdp_sim}. This overestimation does not sacrifice the quality of the decoded videos.}

\subsection{Finite State Problem Formulation}
\label{sec:finiteformulation}
\textcolor{black}{Since the state space $\mathcal{V}^{\mathrm{post}}$ is infinite, the state space $\mathcal S$ is also infinite. Optimizing the scheduling policy over this infinite-state space is intractable. We define a set ${\mathcal W}^\mathrm{buf}$ as the data units in window $\mathcal W$ and their associated predictors in ${\mathcal W}^\mathrm{pre}$. With Assumption \ref{cond:window}, the scheduling policy is actually fixed when are all the data in ${\mathcal W}^\mathrm{buf}$ is received. We only need to determine the optimal scheduling policy for states where some of the video data in ${\mathcal W}^\mathrm{buf}$ has not been received, which is a finite state set. The system state, however, still evolves in the infinite state space $\mathcal S$. In the following, we show how to simplify this infinite state space problem to a finite-state problem.}

We define the set of states where some of the video data in ${\mathcal W}^\mathrm{buf}$ has not been received as follows:
\begin{eqnarray}
{\mathcal S}_\mathrm{W}=\left\{{\mathbf s}| {\mathbf s}\in{\mathcal S},~{\mathcal W}^\mathrm{buf}\not\subset{\mathcal O}({\mathbf s})\right\},
\end{eqnarray}
where ${\mathcal O}({\mathbf s})$ is the set of buffered video data units when the state is ${\mathbf s}$. We define another subset of $\mathcal S$ as the complement of ${\mathcal S}_\mathrm{W}$:
\begin{equation}
{\mathcal S}_\mathrm{\overline W}=\left\{{\mathbf s}|{\mathbf s}\in{\mathcal S},{\mathcal W}^\mathrm{buf}\subseteq{\mathcal O}({\mathbf s})\right\}.
\end{equation}
For all the states in ${\mathcal S}_\mathrm{\overline W}$, all the video data units in ${\mathcal W}^\mathrm{buf}$ has been received.

Given a policy $\mu(\cdot)$, the system state evolves as a controlled Markov chain in set $\mathcal S_\mathrm{W}\cup{\mathcal S}_\mathrm{\overline W}$. Because the transmission rate is finite, the number of states in ${\mathcal S}_\mathrm{\overline W}$ which can be reached from $\mathcal S_\mathrm{W}$ in one step is also finite. We formally define this set of states as follows
\begin{equation}
{\mathcal S_\Delta}=\{{\mathbf s'}|{\mathbf s}'\in {\mathcal S}_\mathrm{\overline W};~\exists\ {\mathbf s}\in \mathcal S_\mathrm{W},~s.t.,~{\mathbb P_{\mu}}({\mathbf s}'|{\mathbf s})>0 \},
\end{equation}
where $\mathbb P_{\mu}({\mathbf s}'|{\mathbf s})$ is the state transition probability under policy $\mu$ (for the expression for $\mathbb P_{\mu}({\mathbf s}'|{\mathbf s})$, see Appendix. \ref{sec:transition}). Thus to move from $\mathcal S_\mathrm{W}$ into the set $\mathcal S_\mathrm{\overline W}$, the system state first hits a state in $\mathcal S_{\Delta}$ and then stays in $\mathcal S_\mathrm{\overline W}$ for some time. During this period, the decoded video distortion is always $d_L$, because all the layers in ${\mathcal W}^\mathrm{buf}$ are available. The evolution of the system when it moves into set ${\mathcal S}_\mathrm{\overline W}$ affect the performance of the system. In general, the longer it stays in ${\mathcal S}_\mathrm{\overline W}$, the better the performance is. Although the scheduling policy in ${\mathcal S}_\mathrm{\overline W}$ is fixed as described in Assumption \ref{cond:window}, the policy in $\mathcal S_\mathrm{W}$ determines how frequently the system state will hit ${\mathcal S}_\mathrm{\overline W}$ and thus critically impacts the system performance.

\begin{figure}[ht]
\centering
\subfigure[$\Pi_\mu$]{
\includegraphics[scale=0.22]{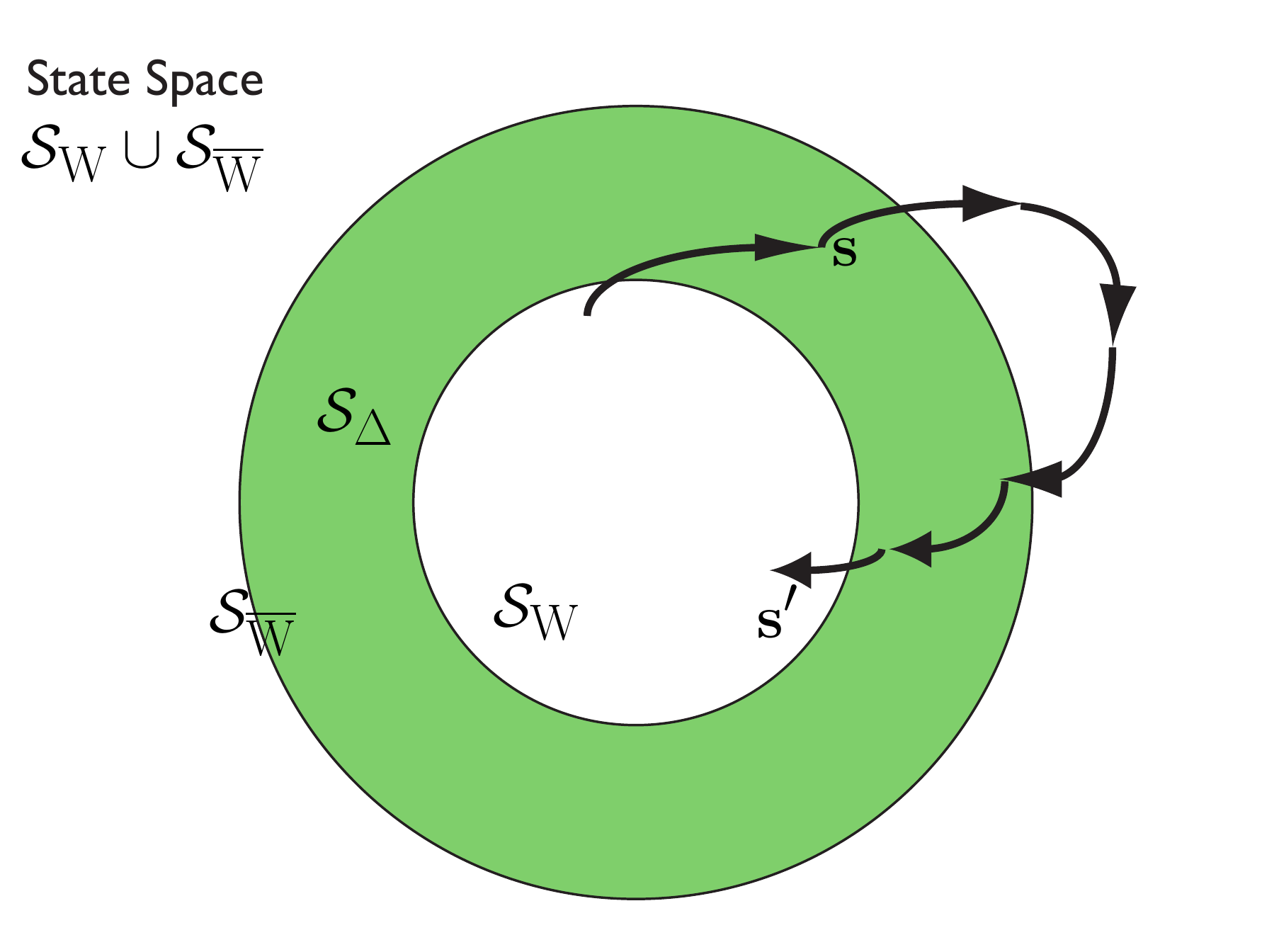}
\label{fig:statespace_A}
}
\subfigure[$\widetilde \Pi_\mu$]{
\includegraphics[scale=0.22]{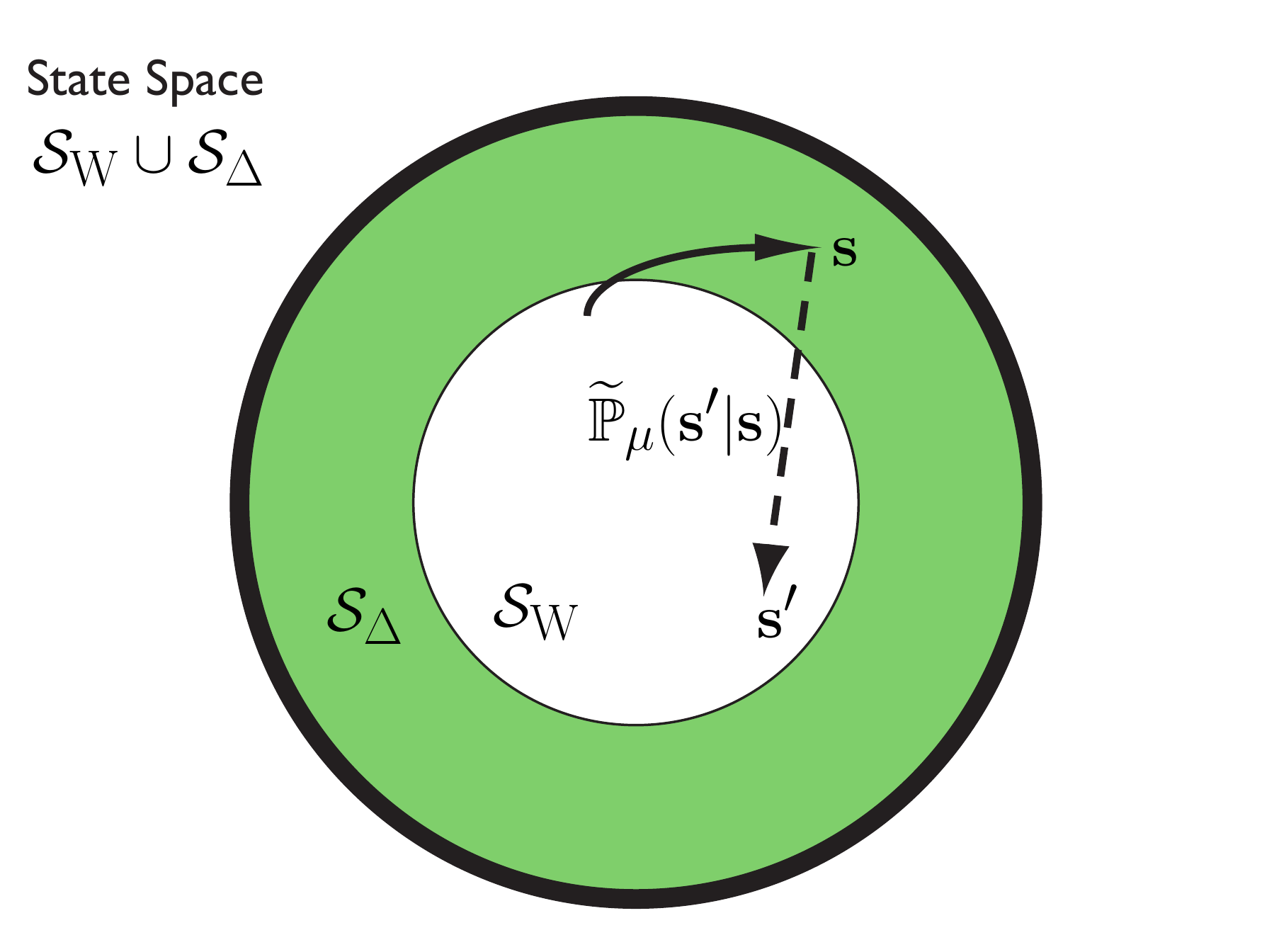}
\label{fig:statespace_simA}
}
\caption[Optional caption for list of figures]{
The dynamics of the system $\Pi_\mu$ and the corresponding simplified system $\widetilde \Pi_\mu$.
}
\label{fig:statespace}
\end{figure}

In the following, we denote the system under a given policy $\mu$ as system $\Pi_{\mu}$. Let $t_\mu({\mathbf s})$ be the expected time spent by $\Pi_\mu$ in ${\mathcal S}_\mathrm{\overline W}$ after it enters ${\mathcal S}_\mathrm{\overline W}$ at state ${\mathbf s}\in \mathcal{S}_{\Delta}$. Let $\widetilde{\mathbb P}_\mu({\mathbf s}'|{\mathbf s})$ denote the probability that $\Pi_\mu$ jumps back to $\mathcal S_\mathrm{W}$ at state ${\mathbf s}'\in\mathcal S_\mathrm{W}$ after it enters ${\mathcal S}_\mathrm{\overline W}$ at state ${\mathbf s}$. To find the optimal policy, we define a finite-state system $\widetilde \Pi_{\mu}$ as follows
\newtheorem{mydef}{\bf Definition}
\begin{mydef}
A system $\widetilde \Pi_\mu$ is called the simplified system of the original system $\Pi_\mu$ if it has the following dynamics:
\begin{enumerate}
\item The system is a controlled semi-Markov process over state space $\mathcal{\widetilde S}=\mathcal{S}_\mathrm{W}\cup\mathcal{S}_\Delta$. In any state ${\mathbf s}\in \mathcal{\widetilde S}$, the distortion is $\mathrm{d}({\mathbf s})$ as in \eqref{eq:key_quality} and \eqref{eq:B_quality}. In any state in $\mathcal{S}_\mathrm{W}$, the system evolves according to the policy $\mu$. The system state transition probability is ${\mathbb P}_\mu(\cdot|\cdot)$.
\item When the system jumps to a state ${\mathbf s}\in \mathcal{S}_\Delta$, it spends $t_\mu({\mathbf s})$ slots in ${\mathbf s}$ with distortion $d_L$ for each slot. The system then transitions to a state ${\mathbf s}'\in \mathcal S_\mathrm{W}$ with probability $\widetilde{\mathbb P}_\mu({\mathbf s}'|{\mathbf s})$ (see Fig. \ref{fig:statespace}).
\end{enumerate}
\end{mydef}

It should be noted that $\widetilde \Pi_\mu$ is not coupled with the original system $\Pi_\mu$. It just shares some properties with the original system. The following theorem relates the distortion under $\widetilde \Pi_\mu$ and that of $\Pi_\mu$.
\newtheorem{thm1}{Theorem}
\begin{thm1}
\label{thm:thm1}
If the jump chain of the original system $\Pi_\mu$ is positive recurrent, then the time-average video distortion of $\Pi_\mu$ is the same as the simplified system $\widetilde \Pi_\mu$.
\end{thm1}
\begin{IEEEproof}[Proof Sketch]
If the jump chain is positive recurrent, the jump from ${\mathcal S}_\mathrm{W}$ to $\mathcal S_\Delta$ can partition the Markov process into i.i.d segments. We only need to optimize the policy $\mu$ to minimize the average distortion in each segment. Every segment consists of two consecutive subsegments. During the first subsegment, ${\mathbf s}\in{\mathcal S_\mathrm{\overline W}}$. In the other subsegment, ${\mathbf s}\in{\mathcal S}_\mathrm{W}$. Because every state in ${\mathcal S}_\mathrm{\overline W}$ has the same distortion $d_L$, we can abstract the first subsegment as a single state with transition probability $\widetilde{\mathbb P}_\mu(\cdot|\cdot)$. This simplified system provides the same average distortion as the original system. For a detailed proof, see the technical report \cite{report}.% \ref{sec:proof}.
\end{IEEEproof}

\begin{remark}
The positive recurrent condition for the jump chain means that the average throughput of the channel is neither too large nor too small relative to the average data rate of the video. If the average throughput of the channel is very large, the receiver buffer can always buffer enough frames and dynamic scheduling is unnecessary. If the average channel throughput is too small, the channel cannot support the video stream and dynamic scheduling cannot help either.
\end{remark}
As indicated by Theorem \ref{thm:thm1}, given any policy $\mu$, the video distortion of $\Pi_\mu$ is the same as $\widetilde \Pi_\mu$. Thus, we can optimize our policy with respect to $\widetilde \Pi_\mu$, which has a finite-state space, and a standard policy optimization algorithm can by applied.

Before we can apply an MDP algorithm to optimize the policy, we need to compute $t_\mu({\mathbf s})$ and $\widetilde{\mathbb P}_\mu({\mathbf s}'|{\mathbf s})$ for every state ${\mathbf s}\in\mathcal {S}_\Delta$ and ${\mathbf s}'\in\mathcal S_\mathrm{W}$. Both $t_\mu({\mathbf s})$ and $\widetilde{\mathbb P}_\mu({\mathbf s}'|{\mathbf s})$ only involve dynamics of the system in $\mathcal S_\mathrm{\overline W}$. Details on how to compute $t_\mu({\mathbf s})$ and $\widetilde{\mathbb P}_\mu({\mathbf s}'|{\mathbf s})$ are found in \cite{report}.% Appendix \ref{sec:computation}.

\subsection{Determining Optimal Policy via Value Iteration}
Given $t_\mu(\cdot)$ and $\widetilde{\mathbb P}_\mu(\cdot|\cdot)$, the optimal policy for an MDP can be determined for the simplified system $\widetilde{\Pi}_\mu$, which is also the optimal policy of $\Pi_{\mu}$. Let ${\mathbf s}^{\mathrm{ini}}$ be any state in $\mathcal{\tilde S}=\mathcal{S}_\mathrm{W}\cup\mathcal{S}_\Delta$. The hitting time to state ${\mathbf s}^{\mathrm{ini}}$ can partition the process into i.i.d cycles. Optimizing the policy $\mu(\cdot)$ in the cycles minimizes the time-average video distortion of the system. Similar to the derivation in \cite[p.~441]{MDP}, this is equivalent to an average-cost minimization problem with stage-cost $\left(\mathrm{d}({\mathbf s})-\lambda\right)\eta({\mathbf s})$, where $\lambda$ is the expected average-cost of each cycle, i.e., the average distortion. The function $\eta(\mathbf{s})$ is defined as
\begin{equation*}
\eta({\mathbf s})=\left\{
\begin{array}{lcr}
1&: &{\mathbf s}\in{\mathcal S_\mathrm{W}}\\
t_\mu({\mathbf s})& :& {\mathbf s}\in{{\mathcal S}_\Delta},	
\end{array}
\right.
\end{equation*}
\textcolor{black}{Note that $\mathrm{d}(\mathbf{s})$ is the cost of spending one slot on state $\mathbf{s}$ and $\lambda$ is the expected cost per slot. Therefore, $\mathrm{d}(\mathbf{s})-\lambda$ is the extra-cost of spending one slot on state $\mathbf{s}$. Since $\eta(\mathbf{s})$ is the average time spent on state $\mathbf{s}$, $(\mathrm{d}(\mathbf{s})-\lambda)\eta({\mathbf{s}})$ is the total extra-cost of visiting state $\mathbf{s}$.} Let us denote by $\mathrm{h}({\mathbf s})$ the average cost-to-go in each cycle when the system starts at state ${\mathbf s}$. Then we have the following Bellman's equation array:
\begin{align}
\label{eq:bellman}
\mathrm{h}({\mathbf s})=
\begin{cases}
\left(\mathrm{d}({\mathbf s})-\lambda\right)\eta({\mathbf s})+ \sum\limits_{\mathbf{s}' \in {\mathcal{S}_\mathrm{W}} \cup {\mathcal S_\Delta}} {{\mathbb P_\mu({\mathbf s}'|{\mathbf s})}\mathrm{h}({\mathbf s}')},&\text{if }\mathbf{s}\in\mathcal{S}_\mathrm{W}\\
\left(\mathrm{d}({\mathbf s})-\lambda\right)\eta({\mathbf s})+ \sum\limits_{\mathbf{s}' \in {S_\mathrm{W}} \cup {\mathcal S_\Delta}} {{{\tilde{\mathbb P}}_\mu({\mathbf s}'|{\mathbf s})}\mathrm{h}({\mathbf s}')},&\text{if }\mathbf{s}\in\mathcal{S}_\Delta\\
\end{cases}
\end{align}
where $\mathrm{h}({\mathbf s}^{\mathrm{ini}}) = 0$. To find the optimal policy, the standard value iteration algorithm can be applied \cite[p.~430]{MDP}.

On the one hand, the assumptions on scheduling policy result in the finite state MDP-based formulation. On the other hand, the assumptions may render the derived scheduling policy sub-optimal. To verify the performance of the scheduling policy derived from the MDP formulation is actually close to optimal, we prove a performance upper bound in the next section.

\subsection{Performance Upper Bound}
\textcolor{black}{
As discussed in Section. \ref{sec:RDmodel}, $\left\{\mathrm{d}^k(z_t), k\in{\mathcal K}\right\}$ are the rate quality models of type-$k$ frames when all the predictors have also been received. Since $\mathrm{d}^k(z_t)$ does not incorporate the distortion due to drift, the time-average distortion of the transmitted video is at least $\frac{1}{n}\sum_{t=1}^{n}\sum_{k\in\mathcal{K}}\mathrm{d}^k(z_t)\mathbbm 1_t^k$, where $n$ is the number of frames in the video sequence and $\mathbbm 1_t^k$ is the indicator that the $t^\mathrm{th}$ frame is a type-$k$ frame. Let $r_t$ be the amount of data that is received in the $t^\mathrm{th}$ slot, a distortion lower bound of any scheduler is given by the following offline optimization problem:
\begin{equation}
\begin{split}
\underset{z_{1:n}}{\mathrm{minimize}} &\quad\frac{1}{n}\sum_{k\in\mathcal{K}}\sum_{t=1}^{n}\mathrm{d}^k(z_t)\mathbbm 1_t^k\\
\mathrm{s.t.}    &\quad\frac{1}{t}\sum_{i=1}^t z_{i}\leq\frac{1}{t}\sum_{i=1}^t r_{i},\quad \forall t\in\{1,2,\cdots,n\},
\end{split}
\label{eq:relax1}
\end{equation}
where the constraint $\frac{1}{t}\sum_{i=1}^t z_{i}\leq\frac{1}{t}\sum_{i=1}^t r_{i}$ guarantees that the received data for the frames displayed before time $t$ does not exceed the cumulative throughput prior to time $t$. We can further relax the constraints in \eqref{eq:relax1} by only keeping the last one, i.e., when $t=n$. The relaxed optimization problem is then given by
\begin{equation}
\begin{split}
\underset{z_{1:n}}{\mathrm{minimize}} & \quad\frac{1}{n}\sum_{k\in\mathcal{K}}\sum_{t=1}^{n}\mathrm{d}^k(z_t)\mathbbm 1_t^k\\
\mathrm{s.t.}    & \quad\frac{1}{n}\sum_{t=1}^n z_t\leq\frac{1}{n}\sum_{t=1}^n r_t.
\end{split}
\label{eq:relax2}
\end{equation}
Let $\widehat{\mathrm{d}^k}(z_t)$ be the convex envelope of $\mathrm{d}^k(z_t)$ (see Fig. \ref{fig:RD}). Since, $\mathrm{d}^k(z_t)$ are lower bounded by $\widehat{\mathrm{d}^k}(z_t)$, we can bound problem \eqref{eq:relax2} by:
\begin{equation}
\begin{split}
\underset{z_{1:n}}{\mathrm{minimize}} & \quad\frac{1}{n}\sum_{k\in\mathcal{K}}\sum_{t=1}^{n}\widehat{\mathrm{d}^k}(z_t)\mathbbm 1_t^k\\
\mathrm{s.t.}    & \quad\frac{1}{n}\sum_{t=1}^n z_t\leq\frac{1}{n}\sum_{t=1}^n r_t.
\end{split}
\label{eq:relax3}
\end{equation}
Let $n^k=\sum_{t=1}^n{\mathbbm 1_t^k}$ denotes the number of type-$k$ frames. Since the functions $\widehat{\mathrm{d}^k}(z_t)$ are convex, by Jensen's inequality, we have
\begin{equation*}
\frac{1}{n^k}\sum_{t=1}^{n}\widehat{\mathrm{d}^k}(z_t)\mathbbm 1_t^k\geq\widehat{\mathrm{d}^k}\left(\frac{1}{n^k}\sum_{t=1}^{n}z_t\mathbbm 1_t^k\right)
\end{equation*}
Problem \eqref{eq:relax3} can then be bounded by:
\begin{equation}
\begin{split}
\underset{z_{1:n}}{\mathrm{minimize}} & \quad\sum_{k\in\mathcal{K}}\frac{n^k}{n}\widehat{\mathrm{d}^k}\left(\frac{1}{n^k}\sum_{t=1}^{n}z_t\mathbbm 1_t^k\right)\\
\mathrm{s.t.}    & \quad\sum_{k\in\mathcal{K}}\frac{n^k}{n}\left(\frac{1}{n^k}\sum_{t=1}^{n}z_t\mathbbm 1_t^k\right)\leq\frac{1}{n}\sum_{t=1}^n r_t.
\end{split}
\label{eq:relax4}
\end{equation}
If the video is reasonably long, e.g. several minutes, the frame number $n$ will be very large. If we let $n\to \infty$ and assume the channel throughput $r_t$ is ergodic, $\frac{1}{n}\sum_{t=1}^n r_t$ will converge to the ergodic capacity $r^\mathrm{avg}=\lim_{n\rightarrow \infty}\frac{1}{n}\sum_{t=1}^n r_t$. Furthermore, let $F^k$ denote the number of type-$k$ frames in a intra period. We have $\frac{n^k}{n}\to\frac{F^k}{F^\mathrm{intra}}$. Similarly, for stationary policies\footnote{A policy is called stationary if it is a function of state $\mathbf s$ and the function is invariant with respect to time $t$.}, the limits $z^k=\lim_{n\rightarrow \infty}\frac{1}{n^k}\sum_{t=1}^{n}z_t\mathbbm 1_t^k$ exist. We have $\lim_{n\to \infty}\left[\frac{n^k}{n}\left(\frac{1}{n^k}\sum_{t=1}^n z_t{\mathbbm 1}_t^k\right)\right]=\frac{F^k}{F^\mathrm{intra}}z^k$. Thus, we have shown the following theorem:
\begin{thm1}
\label{thm:thm2}
For ergodic wireless throughput and stationary adaptive scheduling policies the following optimization gives an upper bound on performance (lower bound of distortion):
\begin{equation}
\begin{split}
\underset{z^k,\ k\in{\mathcal K}}{\mathrm{minimize}} & \quad\sum_{k\in\mathcal{K}}\frac{F^k}{F^\mathrm{intra}}\widehat{\mathrm{d}^k}(z^k)\\
\mathrm{s. t.}    & \quad\sum_{k\in\mathcal{K}}\frac{F^k}{F^\mathrm{intra}}z^k\leq r^\mathrm{avg}.
\end{split}
\label{eq:UB}
\end{equation}
\end{thm1}
Since the rate-distortion function $\widehat{\mathrm{d}^k}(\cdot)$ is assumed to be convex, the above optimization problem is convex and easily solved. In Section \ref{sec:mdp_sim}, this performance bound will be employed as a benchmark to evaluate the performance of our MDP-based scheduling policy.}
\subsection{Performance Evaluation of the MDP-based Scheduling Policy}\label{sec:mdp_sim}
In this section, we evaluate the performance of the policy obtained from our MDP-based formulation. The algorithm was evaluated on test sequences ``foreman", ``bus", ``flower",``mobile" and ``Paris" \cite{Testsequence}. \textcolor{black}{These video sequences were encoded using H.264$\slash$SVC reference software JSVM \cite{JSVM} with a base layer and a CGS enhancement layers. The intra-period and IDR period were set to $F^\mathrm{intra}=16$. The GOP length was fixed at $F^\mathrm{GOP}=4$. The quantization parameter (QP) of the base layer, denoted by $\mathrm{QP^{base}}$, were chosen such that the data rate of the base layer is lower than the average channel throughput. The QP of the CGS enhancement layer is set as $\mathrm{QP^{base}}-10$. We employ this configuration to make sure that the channel is at least good enough to support the base layer. Otherwise, any scheduling policy cannot provide acceptable video quality. The CGS is split into two MGS layers. The first MGS layer contains 6 of the 16 transform coefficients of the CGS layer. The other 10 coefficient belongs to the second MGS layer. The QPs and rate-distortion model parameters of the encoded video sequences are shown in table \ref{tab:parameter}. Parallel to \cite{Mehaela10} and \cite{Fu10}, we employ the FSMC channel model proposed in \cite{ZhangKassam99} to model the dynamics of Rayleigh fading channels. The SNR at the receiver is partitioned into 4 regions using the algorithm proposed in \cite{ZhangKassam99}. In our simulations, we set the average SNR to $\Lambda^\mathrm{avg}=10$dB. For each sequence, 200 transmissions were sent over the simulated channel. A startup delay constraint was fixed to $200$ms, i.e., video playback began 6 frames after the transmission began. After each transmission, a trace file that recorded the packet loss in each time slot was generated. We used the bitstream extractor of JSVM to remove those dropped packets. The extracted bitstreams were decoded using the JSVM decoder with frame copy error concealment.} For more details about the FSMC channel model, see Appendix \ref{sec:settings}.

\begin{center}
\begin{table*}[ht]
\renewcommand{\arraystretch}{1.3}
\caption{The encoding parameters and rate-distortion model parameters of the tested sequences.}
\label{tab:parameter}
\centering
\begin{tabular}{c||c|c|c||c|c||c|c}
\hline
\multirow{2}{*}{sequences}&\multicolumn{3}{c||}{layer 0 (base layer)}&\multicolumn{2}{c||}{Layer 1}&\multicolumn{2}{c}{Layer 2}\\
\cline{2-8}
&QP&$\omega^\mathrm{I}_0,\omega^\mathrm{P}_0,\omega^\mathrm{B^1}_0,\omega^\mathrm{B^2}_0$/Byte&$d_0$/MSE&$\omega^\mathrm{I}_1,\omega^\mathrm{P}_1,\omega^\mathrm{B^1}_1,\omega^\mathrm{B^2}_1$/Byte&$d_1$/MSE&$\omega^\mathrm{I}_2,\omega^\mathrm{P}_2,\omega^\mathrm{B^1}_2,\omega^\mathrm{B^2}_2$/Byte&$d_2$/MSE\\
\hline
foreman&30&6712, 2499, 928, 520&16.27&8302, 8293, 3373, 2775&5.491&5844, 5773, 2177, 1893&4.124\\
\hline
bus&38&5920, 2417, 889, 568&100.8&7837, 8003, 3390, 2925&41.35&4636, 4412, 1577, 1339&21.65\\
\hline
flower&40&8261, 2076, 548, 324&172.1&6786, 6900, 1951, 1611&96.66&6633, 6610, 2008, 1545&30.85\\
\hline
mobile&40&9648, 1556, 510, 262&186.0&9090, 9193, 2541, 2171&89.90&7627, 6894, 1973, 1701&37.35\\
\hline
Paris&32&12353, 2640, 865, 463&32.33&9850, 9457, 2103, 1571&18.59&8091, 7987, 2024, 1555&5.420\\
\hline
\end{tabular}
\end{table*}
\end{center}

The performance of the MDP-based scheduling algorithm was tested over the simulated Markov channel models with different Doppler frequencies ($f^\mathrm{d}=5$Hz and $3$Hz, respectively). The simulation results are summarized in Table \ref{tab:nearoptimal} and Table \ref{tab:nearoptimal_2}. The visual quality is measured via the MS-SSIM index which correlates well with human objective judgments \cite{wirelessVQA}. The time-averaged MS-SSIM value is further converted to Difference Mean Opinion Score (DMOS) using the following mapping
\begin{equation}
{q^\mathrm{dmos} = 13.3442\log(1-q^\mathrm{ssim})+3.6226(1-q^\mathrm{ssim})+77.0117},
\label{eq:dmos}
\end{equation}
where $q^\mathrm{ssim}$ denotes the time averaged quality measured in MS-SSIM and $q^\mathrm{dmos}$ is the corresponding DMOS value. Equation \eqref{eq:dmos} is obtained by logistic regression using the MS-SSIM indices and MOS values of the images in the LIVE database \cite{LIVE}. DMOS ranges from $0$ to $100$. Value $0$ means perfect visual quality and value $100$ means bad visual quality. Roughly speaking, value $50$ means fair quality. It can be seen from Table \ref{tab:nearoptimal} and Table \ref{tab:nearoptimal_2} that the DMOS value of the MDP-based scheduling policy is worse than the performance bound by at most $2$, which is visually insignificant. Given that the bound given by Theorem \ref{thm:thm2} is an upper bound (i.e. a lower bound of DMOS value), the MDP-based scheduling policy is indeed near-optimal.

\begin{table}[ht]
\renewcommand{\arraystretch}{1.3}
\caption{The performance of the near-optimal policy in SSIM-predicted DMOS. $f^\mathrm{d}=5$.}
\label{tab:nearoptimal}
\centering
\begin{tabular}{l||c|c|c|c|c}
\hline
&Paris&mobile&flower&bus&foreman\\
\hline
MDP Policy&26.9020&38.9033&34.5826&41.8721&32.2426\\
\hline
Upper bound&25.6017&38.0842&34.0626&41.4600&31.6807\\
\hline
\end{tabular}
\end{table}

\begin{table}[ht]
\renewcommand{\arraystretch}{1.3}
\caption{The performance of the near-optimal policy in SSIM-predicted DMOS. $f^\mathrm{d}=3$.}
\label{tab:nearoptimal_2}
\centering
\begin{tabular}{l||c|c|c|c|c}
\hline
&Paris&mobile&flower&bus&foreman\\
\hline
MDP Policy&27.1376&39.0452&35.7828&42.0808&32.4431\\
\hline
Upper bound&25.2314&37.9431&33.7052&41.2611&31.4852\\
\hline
\end{tabular}
\end{table}

\section{Near-optimal Heuristic On-line Scheduling Algorithm}
\label{sec:nearoptimal}

Although the MDP-based formulation makes it possible to compute a good scheduling policy using value iteration algorithm, off-line computation of such policies requires \emph{a priori} knowledge of the channel dynamics. This motivates us to design a simple on-line scheduling policy that delivers similar performance as the MDP-based policy that only requires little \emph{a priori} knowledge about the channel dynamics.

A good online video scheduling algorithm should explicitly take advantage of the channel dynamics and schedule data from different quality layers as a function of the receiver buffer state. There are three fundamental questions in desiging such a scheduler: 1) How should one incorporate limited knowledge of channel dynamics in adaptive scheduling? 2) How should one determine the number of enhancement layers to schedule? 3) How should one allocate appropriate transmission rate among current and future intra periods? In the following, we will show how to address these fundamental problems by reasonably simplifying the MDP-based scheduling algorithm.

\subsection{Channel Model Simplification}
In a practical wireless communication environment, accurate channel dynamics models such as the state transition probability ${\bf P}^\mathrm{c}$ are not generally available. Some basic characteristics for the channel dynamics can, however, be easily used. At any slot $t$, the instantaneous channel throughput $r_t$ can be estimated using receiver channel state information as \[\hat{r_t}=x_t(1-y_t),\]where $(x_t,y_t)$ is the channel state at $t$ (see Section.~\ref{sec:simplestate}). The ergodic channel throughput $r^\mathrm{avg}$ can be estimated by averaging $\hat{r_t}$ over time. If we model $r_t$ as the realization of a random process $\{R_t, t\in\mathbb{N}\}$, the temporal correlation coefficient $\rho=\frac{\mathrm{cov}(R_t,R_{t+1})}{\sigma(R_t)\sigma(R_{t+1})}$ can also be estimated from $\hat{r_t}$. Further it is reasonable to assume the channel throughput $R_t$ will typically regress to the mean $r^\mathrm{avg}$. This inspires us to use a simple autoregressive model to capture the dynamics of the channel. A first order autoregressive model (AR(1)) for $R_t$ is given as,
\begin{equation}
\label{eq:AR}
R_t-\phi R_{t-1}=c+N_t,
\end{equation}
where $N_t$ is an i.i.d random variable with zero mean value. From \eqref{eq:AR}, parameter $c$ and $\phi$ can be estimated as $\phi=\rho$ and $c=r^\mathrm{avg}(1-\rho)$ \cite{TimeSer}[p.~115]. Thus, we have
\begin{equation}
\label{eq:trueAR}
R_t-\rho R_{t-1}=r^\mathrm{avg}(1-\rho)+N_t.
\end{equation}

Using this autoregressive model, the amount of data that will be delivered in the next $\zeta$ slots by the channel can be estimated as
\begin{equation}
\label{eq:capacity}
\mathrm{g}(\hat{r_t})=\mathbb E\left[\sum_{{a}=0}^{\zeta-1}R_{t+a} \bigg\vert R_t=\hat{r_t}\right]=\sum_{a=0}^{\zeta-1}[\hat{r_t}\rho^{a}+r^\mathrm{avg}(1-\rho^{a})].
\end{equation}
To obtain an accurate estimate in the near future, we set the length of the window $\zeta$ into the future that will be considered to be the relaxation time\footnote{The relaxation time is defined as the temporal distance at which the temporal correlation coefficient is reduced to $\frac{1}{e}$} of the channel, i.e. , $\zeta=\lceil-(\ln\rho)^{-1}\rceil$. In the following, we use this to determine which quality layers to schedule.

\subsection{Layer Selection}

Given the current channel state, receiver buffer state, and estimated available capacity for a window $\zeta$ into the future, the goal is to determine which layers to schedule. We will focus on determining the number of enhancement layers which should be scheduled. We denote by $\mathrm{L}^\mathrm{sch}({\mathbf s}_t)$ the number of layers to be scheduled if the state is ${\mathbf s}_t$. Once $\mathrm{L}^\mathrm{sch}({\mathbf s}_t)$ is determined, the online scheduling algorithm only schedules data units from the first $\mathrm{L}^\mathrm{sch}({\mathbf s}_t)$ layers.

\textcolor{black}{The layer selection scheme for our proposed on-line algorithm is motivated by that of the MDP-based policy. Using $\mathrm{g}(\hat{r_t})$ defined in \eqref{eq:capacity}, we can estimate the amount of data which can be delivered in the next $\zeta$ slots. Let $\Gamma({\ell},{\mathbf s}_t)$ be the amount of data which is not currently available at the playback buffer at time $t$, and belongs to the first $\ell$ layers of the next $\zeta$ frames that have not been decoded. The quantities $\mathrm{g}(\hat{r_t})$ and $\Gamma({\ell},{\mathbf s}_t)$ summarize the channel and buffer states for the next $\zeta$ slots. Note that $\Gamma({\ell-1},{\mathbf s}_t)\leq \mathrm{g}(\hat{r_t})<\Gamma({\ell},{\mathbf s}_t)$ means that we can probably transmit all the data up to the $\ell^\mathrm{th}$ layer in the next $\zeta$ slots. Intuitively, we can simply choose $\mathrm{L}^\mathrm{sch}({\mathbf s}_t)=\ell-1$ when $\Gamma({\ell-1},{\mathbf s}_t)\leq \mathrm{g}(\hat{r_t})<\Gamma({\ell},{\mathbf s}_t)$. As discussed next, this layer selection scheme can be motivated by the near-optimal scheduling policies computed for the MDP-based model.}

\begin{figure}[ht]
\centering
\subfigure[$f^\mathrm{d}=5Hz$.]{
\includegraphics[scale=0.35]{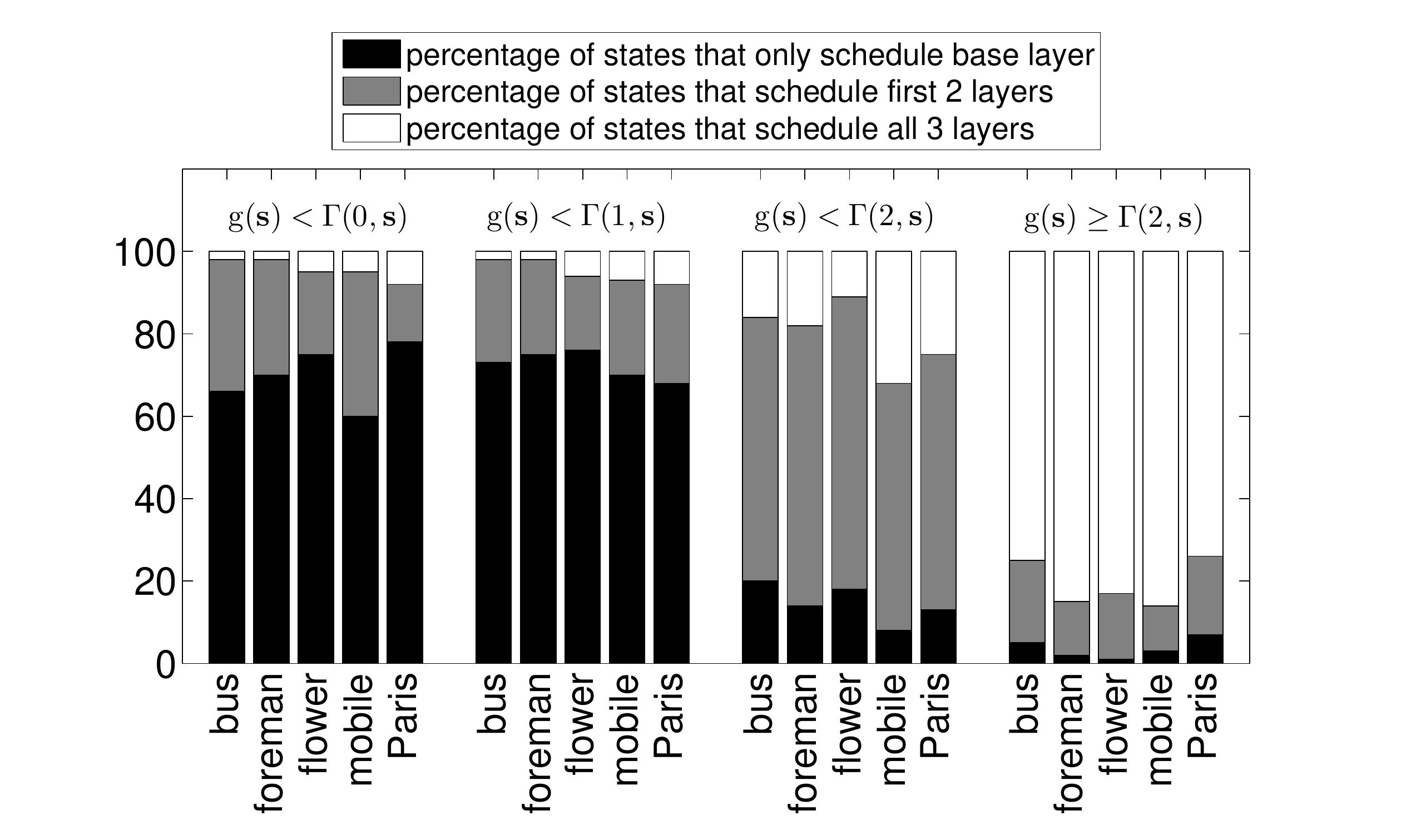}
}
\subfigure[$f^\mathrm{d}=3Hz$.]{
\includegraphics[scale=0.35]{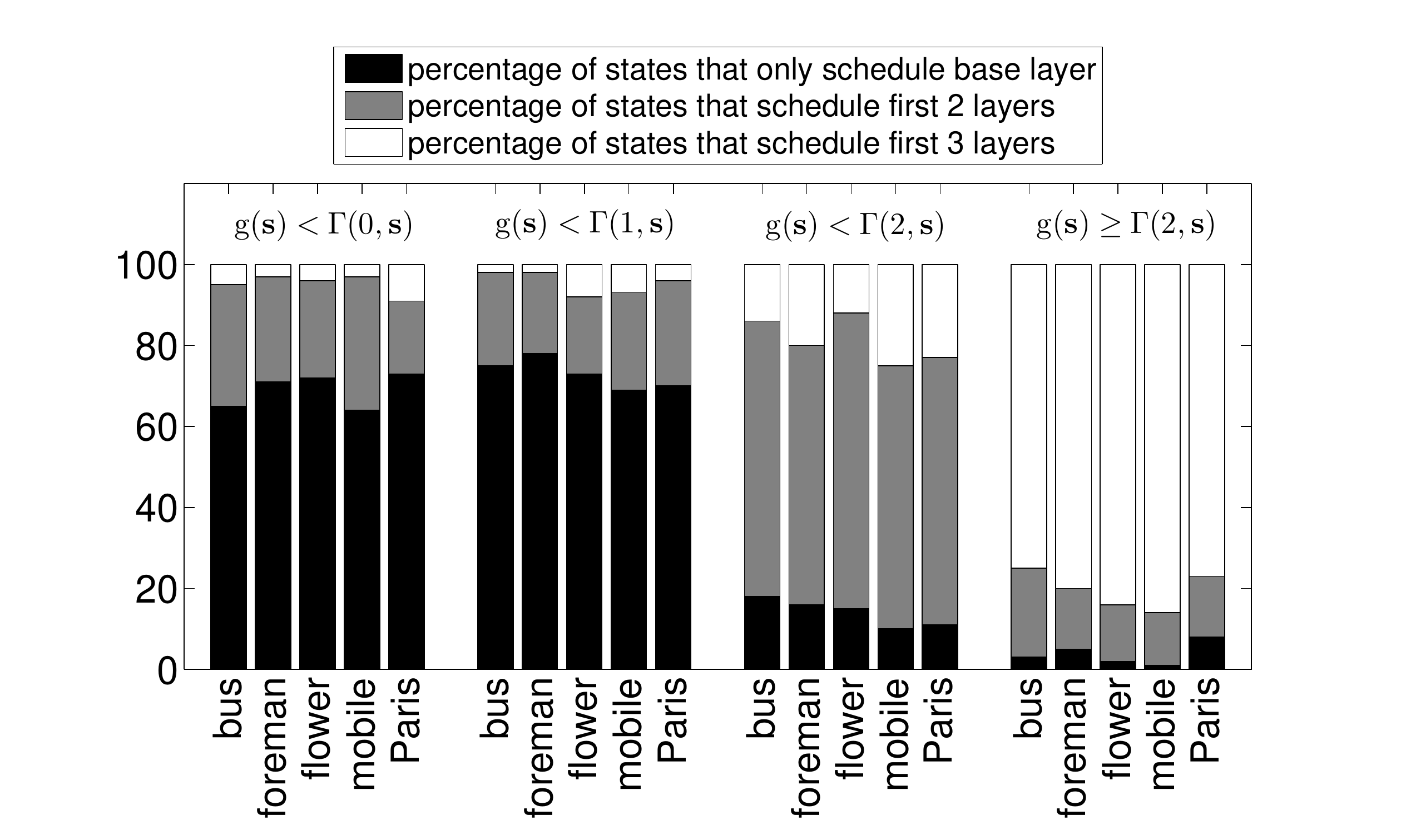}
}
\caption[Optional caption for list of figures]{
Given different relationship between $\mathrm{g}({\mathbf s})$ and $\Gamma({\ell},{\mathbf s})$, the proportions of states corresponding to different $\mathrm{L}^\mathrm{sch}({\mathbf s})$ are shown in different colors. Results are obtained under Rayleigh fading channels with different Doppler shifts (5Hz in (a) and 3Hz in (b)) and are calculated on 5 different video sequences (``bus", ``foreman", ``flower", ``mobile", ``Paris").
}
\label{fig:layer choice}
\end{figure}

\textcolor{black}{Note that $\hat{r_t}=x_t(1-y_t)$ is determined by state ${\mathbf s}_t$, thus $\mathrm{g}(\hat{r_t})$ can also be written as function of ${\mathbf s}_t$, i.e., $\mathrm{g}({\mathbf s}_t)$. Suppose we partition the state space into subsets $\mathcal P^\ell=\{{\mathbf s}\in \mathcal S: \Gamma({\ell-1},{\mathbf s})\leq \mathrm{g}(\mathbf{s})<\Gamma({\ell},{\mathbf s})\}, \ell\in\{1,\cdots,L+2\}$ and calculate the fraction of states in $\mathcal P^\ell$ where the MDP-based policy only schedules the first $\ell-1$ layers\footnote{We define $\Gamma({L+2},{\mathbf s}_t)=+\infty$}. As shown in Fig. \ref{fig:layer choice}, for 71\% of the states of $\mathcal P^1$ and $\mathcal P^2$, the MDP-based policy only schedules the first layer. For  about 65\% of the states of $\mathcal P^3$, the MDP-based policy only schedules the first 2 layers. Finally the MDP-based policy will schedule all the layers on 81\% of the states in $\mathcal P^4$. These observations justify our intuition regarding layer selection. In our proposed on-line scheduling algorithm, we will simply choose $\mathrm{L}^\mathrm{sch}({\mathbf s}_t)=\ell-1$ if $\Gamma({\ell-1},{\mathbf s}_t)\leq \mathrm{g}(\hat{r_t})<\Gamma({\ell},{\mathbf s}_t)$. In other words, our heuristic algorithm determines $\mathrm{L}^\mathrm{sch}({\mathbf s}_t)$ by roughly estimating the number of layers which can be transmitted.}

\subsection{Resource Allocation Between Current and Future Intra Periods}
In each transmission slot, about $\hat{r_t}$ bits of video data are delivered to the receiver. In the following, we refer to $\hat{r_t}$ as the budget for slot $t$. Once $\mathrm{L}^\mathrm{sch}({\mathbf s}_t)$ is determined, we still need to determine how to allocate this budget among current and future intra periods. Sometimes it is necessary to transmit data associated with next $I$ frame before the data units in the current intra period. For example, when the next $I$ frame is approaching its display deadline and its base layer has not yet been received, if we focus on transmitting the frames in the current intra period sequentially, this increases the risk that the next $I$ frame can not be decoded before its deadline. This in turn would cause severe decoding failures throughout the next intra period.

Let $\mathcal I$ be the data units in the undecoded I frame that has the earliest display deadline. We denote by $\Psi^\mathrm{cur}(\ell,{\mathbf s}_t)$ the amount of unreceived data in the first $\ell^\mathrm{th}$ layer of current intra period at state ${\mathbf s}_t$. We denote by $\Psi^\mathrm{I}(\ell,{\mathbf s}_t)$ the amount of unreceived data in the first $\ell^\mathrm{th}$ layer of $\mathcal I$ at state ${\mathbf s}_t$. We propose the following heuristic for allocating the bit budget between current intra period and $\mathcal I$. In each transmission slot, the scheduling algorithm allocates up to $\Omega_t=\frac{\Psi^\mathrm{I}(\mathrm{L}^\mathrm{sch}({\mathbf s}_t),{\mathbf s}_t)}{\Psi^\mathrm{cur}(\mathrm{L}^\mathrm{sch}({\mathbf s}_t),{\mathbf s}_t)+\Psi^\mathrm{I}(\mathrm{L}^\mathrm{sch}({\mathbf s}_t),{\mathbf s}_t)}$ of the transmission bit budget to $\mathcal I$. In other words, the number of bits allocated to $\mathcal I$ is $\min(\Omega_t\times \hat{r_t},\Psi^\mathrm{I}(\mathrm{L}^\mathrm{sch}({\mathbf s}_t),{\mathbf s}_t))$.

Here $\Omega_t$ gives the relative importance of the next $I$ frame and current intra period. If $\Psi^\mathrm{I}(\mathrm{L}^\mathrm{sch}({\mathbf s}_t),{\mathbf s}_t)=0$, then $\Omega_t=0\%$. It is not necessary to transmit any data for the next $I$ frame. If $\Psi^\mathrm{pre}(\mathrm{L}^\mathrm{sch}({\mathbf s}_t),{\mathbf s}_t)=0$, then $\Omega_t=100\%$. We only focus on transmitting the future intra periods.

The online scheduling algorithm is summarized in Algorithm \ref{alg:online}.
\begin{algorithm}[h]
\caption{On-line adaptive scheduling algorithm}
\label{alg:online}
\begin{algorithmic}[1]
\Require ${\mathbf s}_t$, $r^\mathrm{avg}$, $x_t$, $y_t$, and $\rho$
\State $\zeta=\lceil-(\ln\rho)^{-1}\rceil$; $\hat{r_t}=x_t(1-y_t)$
\Loop{ $t$}
    \State $\mathrm{g}(\hat{r_t}) \gets \sum_{a=0}^{\zeta-1}[\hat{r_t}\rho^{a}+r^\mathrm{avg}(1-\rho^{a})]$
    \For{$\ell = 1 \to L+1$}
        \State Compute $\Gamma({\ell},{\mathbf s}_t)$
        \If{$\mathrm{g}(\hat{r_t})<\Gamma({\ell},{\mathbf s}_t)$}
            \State break
        \EndIf
    \EndFor
    \If{$\ell$=1}
        \State $\mathrm{L}^\mathrm{sch}({\mathbf s}_t)\gets1$
    \Else
        \State $\mathrm{L}^\mathrm{sch}({\mathbf s}_t)\gets\ell-1$
    \EndIf
    \State Compute $\Psi^\mathrm{cur}(\mathrm{L}^\mathrm{sch},{\mathbf s}_t)$ and $\Psi^\mathrm{I}(\mathrm{L}^\mathrm{sch},{\mathbf s}_t)$
    \State $\Omega_t\gets \frac{\Psi^\mathrm{I}(\mathrm{L}^\mathrm{sch},{\mathbf s}_t)}{\Psi^\mathrm{cur}(\mathrm{L}^\mathrm{sch},{\mathbf s}_t)+\Psi^\mathrm{I}(\mathrm{L}^\mathrm{sch},{\mathbf s}_t)}$
    \State Schedule $\min(\Omega_t\times \hat{r_t},\Psi^\mathrm{I}(\mathrm{L}^\mathrm{sch},{\mathbf s}_t))$ bits from $\mathcal I$. \Comment{Scheduling data}
    \State Schedule $\hat{r_t}-\min(\Omega_t\times \hat{r_t},\Psi^\mathrm{I}(\mathrm{L}^\mathrm{sch},{\mathbf s}_t))$ bits from other active frames.
\EndLoop
\end{algorithmic}
\end{algorithm}

\subsection{Performance Evaluation of the On-line Scheduling Algorithm}
\begin{figure}[!t]
\centering
\subfigure[$f^\mathrm{d}=5Hz$.]{
\includegraphics[scale=0.35]{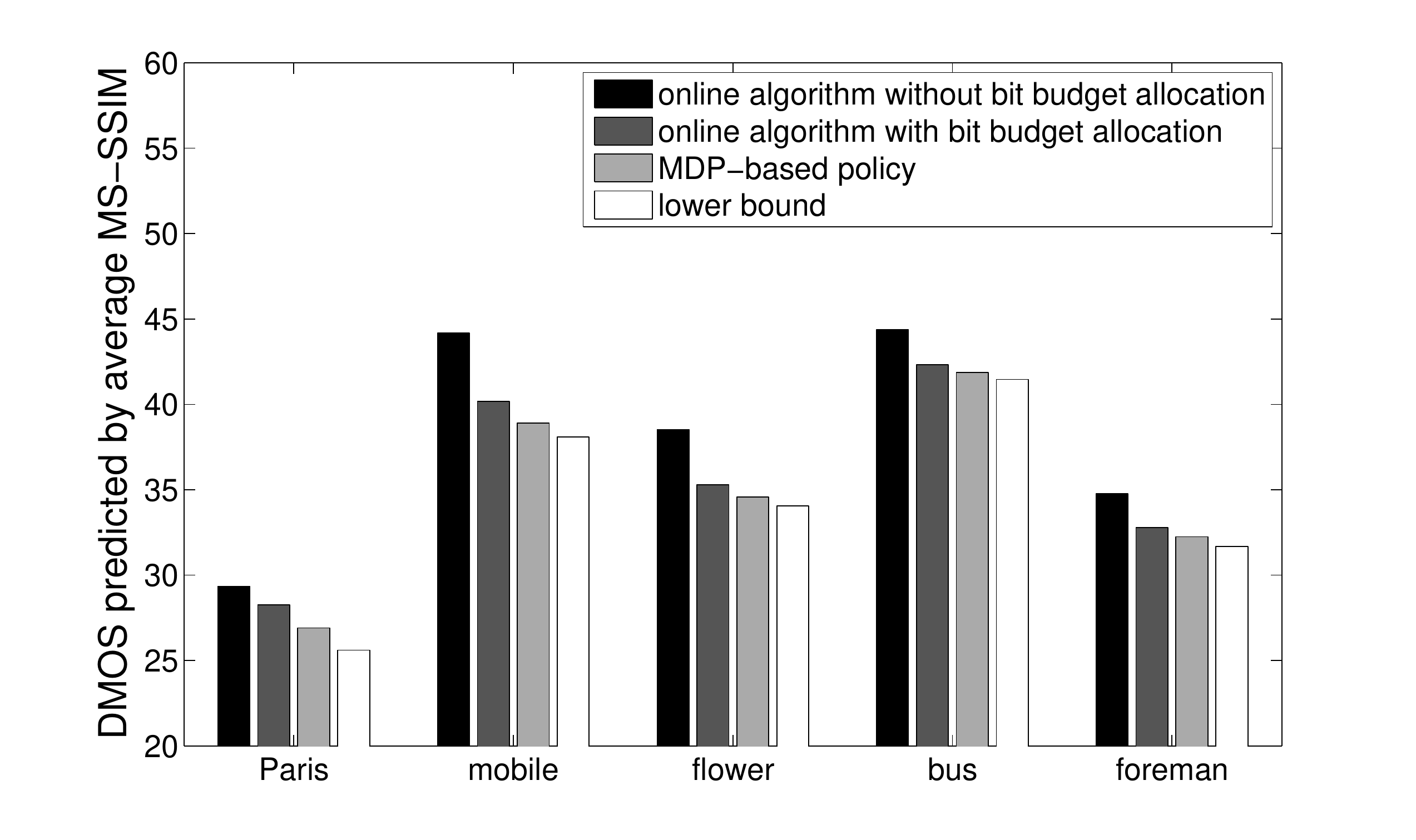}
}
\subfigure[$f^\mathrm{d}=3Hz$.]{
\includegraphics[scale=0.35]{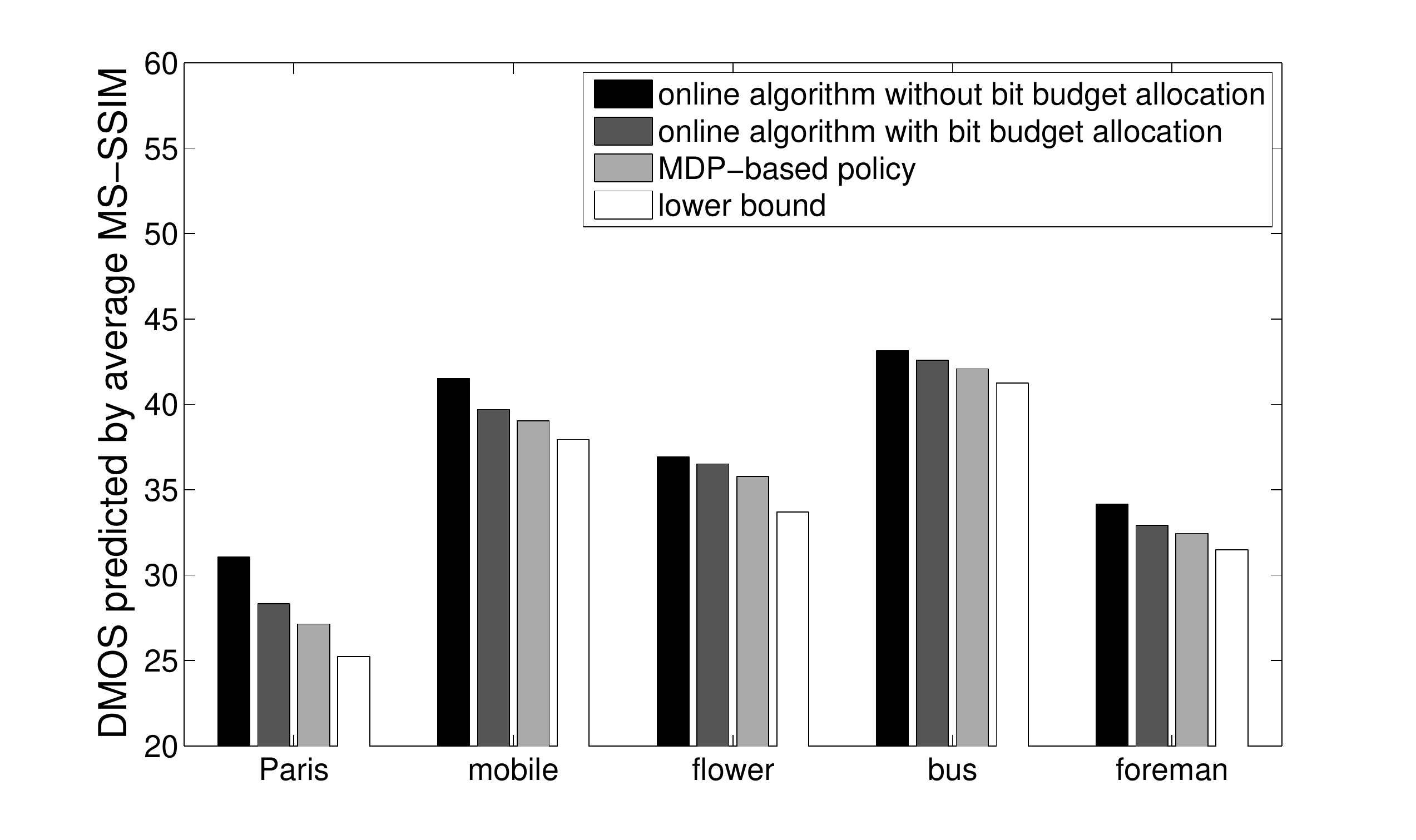}
}
\caption[Optional caption for list of figures]{
Performance comparison of different scheduling algorithms. Video quality is measured in DMOS which is predicted by MS-SSIM using equation \eqref{eq:dmos}.
}
\label{fig:performance}
\end{figure}

The performance of the on-line scheduling algorithm was tested over the simulated Markov channel models with different Doppler frequencies ($f^\mathrm{d}=5$Hz and $3$Hz, respectively). This setting is the same as the simulation setting in Section \ref{sec:mdp_sim}. The results are summarized in Fig. \ref{fig:performance}. As can be seen, the performance of the proposed online-scheduling algorithm is almost as good as the MDP-based scheduling algorithm. Moreover the online scheduling algorithm's performance is close to the bound given by Theorem \ref{thm:thm2}. We conclude it is a near-optimal scheduling algorithm.

We have also tested the performance of the online algorithm without bit budget allocation between current and future intra periods. As can be seen, the performance is worse than the MDP-based scheduling policy and the performance bound. This motivates the necessity of allocating bit between current and future intra periods.
\section{Conclusions}
\label{sec:conclusion}
We have developed adaptive scheduling algorithms for stored scalable video transmission in wireless channels. By modeling the wireless channel as a Markov chain, an MDP model is proposed in which policies that minimize the distortion of decoded videos can be computed. By simplifying the scheduling algorithm obtained from the MDP formulation, we propose an online scheduling algorithm which only requires limited knowledge of channel dynamics. Simulation results demonstrate the near-optimality of the proposed online scheduling policy versus a proposed bound on performance.

\appendices
\section{Transition Probability}\label{sec:transition}
{\it Notations:} Let $\bf 1$ be the unit vector of all-ones and $\bf 0$ be the zero vector. $\max\{\bf a,\bf b\}$ is the componentwise maximum of vector $\bf a$ and $\bf b$. $\mathbbm 1(\cdot)$ is the indicator function.

Let ${\mathbf s}_t=({\mathbf c}_t,{\mathbf v}_t)$ and $\mathcal U_{{\mathbf s}_t}$ be the system state and the corresponding feasible control set at slot $t$, where ${\mathbf c}_t=(x_t,y_t)$ and ${\mathbf v}_t=({v}^{\mathrm{I}}_t,{\mathbf v}^\mathrm{pre}_t, {\mathbf v}^\mathrm{W}_t,{\mathbf v}^\mathrm{post}_t)$. At the beginning of each slot, one frame is decoded and played out. Let ${\mathbf v}^{+}_t=({v}^{\mathrm{I+}}_t,{\mathbf v}^\mathrm{pre+}_t, {\mathbf v}^\mathrm{W+}_t,{\mathbf v}^\mathrm{post+}_t)$ denote the buffer state right after the first frame is displayed.
For ${\mathbf v}^\mathrm{I+}_t$, we have
\begin{equation}
{\mathbf v}^\mathrm{I+}_t=
\begin{cases}
F^\mathrm{intra}-1& \text{if $v^\mathrm{I}_t=0$,}
\\
v^\mathrm{I}_t-1 & \text{if $v^\mathrm{I}_t\neq0$,}
\end{cases}
\end{equation}
The first frame in $\mathcal W$ is moved into ${\mathcal W}^\mathrm{pre}$, thus we have
\begin{equation}
\mathbf{v}_t^\mathrm{pre+}=\left(b^\mathrm{pre}_{f^\mathrm{key}},\cdots,b^\mathrm{pre}_{-1},b^\mathrm{W}_0\right).
\end{equation}
The first frame in ${\mathcal W}^\mathrm{post}$ moves into $\mathcal W$. Thus, we have
\begin{equation}
\label{eq:update3}
\mathbf{v}^\mathrm{W+}_t=\left(b^\mathrm{W}_1,\cdots,b^\mathrm{W}_{W-1},\sum\nolimits_{\ell=0}^L\mathbbm{1}(b_\ell^\mathrm{post}\geq1)\right),
\end{equation}
For the set $\mathcal{W}^\mathrm{post}$, once the current frame is played out, we have
\begin{equation}
\label{eq:update4}
\mathbf{v}^\mathrm{post+}=\max\left\{\mathbf{v}^\mathrm{post}-\mathbf{1},\mathbf{0}\right\}
\end{equation}

After the first frame is displayed, the transmitter begins to sequentially transmit the collection of video data units indicated by the action
${\mathcal U}_t =\mu({\mathbf s}_t)=\{({f_1},{\ell_1}), \cdots, ({f_{|\mathcal U_t|}},{\ell_{|\mathcal U_t|}})\}$. Let $\Delta \mathcal U_t=\{({f_1},{\ell_1}), \cdots, (f_{n_t},{\ell_{n_t}})\}$ denote the completely received data units by the end of the slot, where $n_t$ is the number of received data units. Among the data units in $\Delta \mathcal U_t$, let $\Delta {\mathbf v}^{\mathrm{pre}}_t$ and $\Delta {\mathbf v}^{\mathrm{W}}_t$ be the number of newly received data units for each frame in set ${\mathcal W}^\mathrm{pre+}$ and ${\mathcal W}^\mathrm{W+}$, respectively. At the beginning of the $(t+1)^\mathrm{th}$ slot, we have the following state transition relationship
\begin{eqnarray}
{\mathbf v}^{\mathrm{pre}}_{t+1}&=&{\mathbf v}^{\mathrm{pre+}}_t+\Delta {\mathbf v}^{\mathrm{pre}}_t,\\
{\mathbf v}^{\mathrm{W}}_{t+1}&=&{\mathbf v}^{\mathrm{{W}+}}_t+\Delta {\mathbf v}^{\mathrm{W}}_t.
\end{eqnarray}
Similarly, we denote by $\Delta\mathbf{v}^\mathrm{post}_t=\left(\Delta b^\mathrm{post}_0,\cdots,\Delta {b}^\mathrm{post}_L\right)$ the number of newly received data units for each layer in frame set ${\mathcal W}^\mathrm{post+}$. The state transition relationship of ${\mathcal W}^\mathrm{post}$ is
\begin{equation}
{\mathbf v}^{\mathrm{post}}_{t+1}={\mathbf v}^{\mathrm{post+}}_t+\Delta {\mathbf v}^{\mathrm{post}}_t
\end{equation}
The amount of video data in $\Delta \mathcal U_t$, denoted by $\Phi({\mathbf v}_t,\Delta \mathcal U_t)$, can be estimated according to buffer state $v^\mathrm{I}_t$ and the rate-quality model introduced in Section \ref{sec:RDmodel}. Specifically, for each data unit in $\Delta \mathcal U_t$, we first determine the frame type according to ${v}^\mathrm{I}_t$ and then estimate the amount of data by the rate-quality model. The set $\Delta \mathcal U_t$ records the completely transmitted data units up to $(f_{n_t},\ell_{n_t})^\mathrm{th}$ data unit. However, data unit $(f_{n_t+1},\ell_{n_t+1})$ is only partially received. Denoting the amount of data in unit $(f_{n_t+1},\ell_{n_t+1})$ by $\tilde \Phi({\mathbf v}^\mathrm{I}_t,\Delta \mathcal U_t)$, the amount of received data is at least $\Phi({\mathbf v}^\mathrm{I}_t,\Delta \mathcal U_t)$ and at most $\Phi({\mathbf v}^\mathrm{I}_t,\Delta \mathcal U_t)+\tilde \Phi({\mathbf v}^\mathrm{I}_t,\Delta \mathcal U_t)$. Assuming the physical layer packet length is $L^\mathrm{PHY}$, there is $N=\lceil\frac{x_t}{L^\mathrm{PHY}}\rceil$ packet transmissions during a time slot. The number of successfully transmitted packets is at least $N_\mathrm{l}=\lceil\frac{\Phi({\mathbf v}^\mathrm{I}_t,\Delta \mathcal U_t)}{L^\mathrm{PHY}}\rceil$ and is less than $N_\mathrm{h}=\lceil\frac{\Phi({\mathbf v}^\mathrm{I}_t,\Delta \mathcal U_t)+\tilde \Phi({\mathbf v}^\mathrm{I}_t,\Delta \mathcal U_t)}{L^\mathrm{PHY}}\rceil$. As assumed in Section \ref{sec:channelModel}, the channel state is constant over each slot. Thus, the packet losses are independent within each slot. The number of successful packet transmissions in a slot is distributed binomially. Hence, the state transition probability from ${\mathbf s}_t=({\mathbf c}_t,{\mathbf v}_t)$ to ${\mathbf s}_{t+1}=({\mathbf c}_{t+1},{\mathbf v}_{t+1})$ is
\begin{equation}\label{eq:TranProb}
	\mathbb P_{\mu}({\mathbf s}_{t+1}|{\mathbf s}_t)=\left[\sum_{n_t=N_\mathrm{l}}^{N_\mathrm{h}-1}{{N}\choose{n_t}}y_t^{N-n_t}(1-y_t)^{n_t}\right]\mathbb P({\mathbf c}_{t+1}|{\mathbf c}_t),
\end{equation}
where the first multiplicative term is the transition probability of the receiver buffer state from ${\mathbf v}_t$ to ${\mathbf v}_{t+1}$ and the second term is the transition probability of the channel state from ${\mathbf c}_t$ to ${\mathbf c}_{t+1}$.

\section{Simulation Settings}
\label{sec:settings}
We employ the FSMC channel model proposed in \cite{ZhangKassam99} to model the dynamics of Rayleigh fading channels. The SNR at the receiver is partitioned into $|\mathcal C|$ regions using the algorithm proposed in \cite{ZhangKassam99}. Let $\Lambda_i$ be the partition thresholds, where $\Lambda_0=-\infty$ and $\Lambda_{|\mathcal C|}=\infty$. Let $\tilde \Lambda_{k}$ be the representative SNR in the $k^\mathrm{th}$ region. For Rayleigh fading channels, we have
\begin{equation}
\tilde \Lambda_k=\frac{\int_{\Lambda_{k-1}}^{\Lambda_k}\lambda \mathrm{p}(\lambda)\mathrm{d}\lambda}{\int_{\Lambda_{k-1}}^{\Lambda_{k}} \mathrm{p}(\lambda)\mathrm{d}\lambda},
\end{equation}
where $\mathrm{p}(\lambda)=\frac{1}{\Lambda^\mathrm{avg}}\exp(-\frac{\lambda}{\Lambda^\mathrm{avg}})$ is the probability distribution function of the received instantaneous SNR of Rayleigh fading channels with average SNR $\Lambda^\mathrm{avg}$. According to \cite{ZhangKassam99}, the state transition probability ${\bf P}^c$ is computed as
\begin{equation*}
{\bf P}^c_{i,j}=
\begin{cases}
\frac{\mathcal K(\Lambda_{j})\Delta T}{\pi_i}& \text{if $j=i+1$,}
\\
\frac{\mathcal K(\Lambda_i)\Delta T}{\pi_i}& \text{if $j=i-1$,}
\\
1-\frac{\mathcal K(\Lambda_{j})\Delta T}{\pi_i}-\frac{\mathcal K(\Lambda_i)\Delta T}{\pi_i}&\text{if $j=i$,}
\\ 0 &\mathrm{otherwise,}
\end{cases}
\end{equation*}
where $\pi_i=\int_{\Lambda_{i-1}}^{\Lambda_i} \mathrm{p}(\lambda)\mathrm{d}\lambda$. $\mathcal K(\Lambda_i)=\sqrt{\frac{2\pi \Lambda_i}{\Lambda^\mathrm{avg}}}f^\mathrm{d}\exp(-\frac{\Lambda_i}{\Lambda^\mathrm{avg}})$ is the level crossing rate of threshold $\Lambda_i$ where $f^\mathrm{d}$ is the Doppler frequency. The coherence time is estimated via $t_\mathrm{cor}=0.423/f^\mathrm{d}$. In our simulations, we set $|\mathcal C|=4$ and $\Lambda^\mathrm{avg}=10$dB.

We assume that BPSK, QPSK and 8PSK are used for modulation. The symbol error rate $p^s_{k}$ in the $k^\mathrm{th}$ SNR region is $p^s_k=2Q(\sqrt{2\tilde\Lambda_k}\sin\frac{\pi}{2^M})$, where $M=1,2,3$ for BPSK, QPSK and 8PSK, respectively. Each packet contains 2048 symbols. Thus, the packet length $L^\mathrm{PHY}=2048\times M$, where $M=1,2$ and $3$ for BPSK, QPSK and 8PSK, respectively. The transmission time for each packet is $\Delta t=1.5$ms. The transmission data rate is given by $x_k=\frac{\Delta T}{\Delta t}L^\mathrm{PHY}$. The packet error rate is given by $y_k=1-(1-p^s_k)^{2048}$.  The modulation scheme for $k^\mathrm{th}$ channel states is chosen such that the throughput $x_k(1-y_k)$ is maximized.

\ifCLASSOPTIONcaptionsoff
  \newpage
\fi

%\newpage
\bibliographystyle{IEEEtran}
%\bibliography{IEEEabrv,refs,strings}

\end{document}